%% file: paper.tex
\documentclass[10pt,twocolumn,twoside,a4paper]{article}
\pdfoutput=1
\usepackage{url}
\usepackage{color}
\usepackage{amsmath}
\usepackage{amssymb}
\usepackage{amsfonts}
\usepackage{boxedminipage}
\usepackage{colortbl}
\usepackage{graphicx}
\usepackage[Symbol]{upgreek}

\usepackage[style=alphabetic,sorting=nyt,maxbibnames=10,hyperref=true,backend=bibtex]{biblatex}
\newcommand{\plus}{\ensuremath{\raisebox{0.1em}{$\scriptscriptstyle\mathord{+}$}}}
\renewcommand{\star}{\ensuremath{\raisebox{0.1em}{$\scriptstyle\mathord{*}$}}}

\newcommand{\ccomma}{\raisebox{2pt}{\textbf{,}}}

\makeatletter 
\def\@normalsize{\@setsize\normalsize{10pt}\xpt\@xpt
\abovedisplayskip 10pt plus2pt minus5pt\belowdisplayskip \abovedisplayskip
\abovedisplayshortskip \z@ plus3pt\belowdisplayshortskip 
6pt plus3pt minus3pt\let\@listi\@listI}
\def\subsize{\@setsize\subsize{12pt}\xipt\@xipt}
\def\section{\@startsection
	{section}		
	{1}
	{\z@}
	{2.2ex plus 1ex minus .2ex} 		% {-3.5ex plus-1ex minus -.2ex}
	{1.2ex plus .2ex \@afterindentfalse}	% {2.3ex plus.2ex}
	{\large\bf}}				% {\reset@font\Large\bf}}
\def\subsection{\@startsection
	{subsection}
	{2}
	{\z@}
	{2.0ex plus 1ex}			% {-3.25ex plus-1ex minus-.2ex}
	{.8ex plus .2ex \@afterindentfalse}	% {1.5ex plus.2ex}
	{\subsize\bf}}				% {\reset@font\large\bf}}
\def\subsubsection{\@startsection
	{subsubsection}
	{3}
	{\z@}
	{1.8ex plus 1ex}			% {-3.25ex plus -1ex minus-.2ex}
	{.8ex plus .2ex \@afterindentfalse}	% {1.5ex plus.2ex}
	{\normalsize\bf}}			% {\reset@font\normalsize\bf}}
\def\paragraph{\@startsection 
	{paragraph}
	{4}
	{\z@}
	{1.8ex plus .3ex}			% {3.25ex plus1ex minus.2ex}
	{-1em \@afterindentfalse}		% {-1em}
	{\normalsize\bf}}			% {\reset@font \normalsize\bf}}

\newcommand{\CBGF}{$\Upxi$BGF}

\newenvironment{denselist}{\begin{list}{\textbullet}{\setlength{\itemsep}{0em}\setlength{\parsep}{0em}}}{\end{list}}
\newenvironment{denselista}{\begin{list}{\textbullet}{\setlength{\itemsep}{.2em}\setlength{\parsep}{0em}}}{\end{list}}

\title{\textbf{The Grammar Hammer of 2012\footnote{The title relates both to the folklore story of a steel driving man named John Henry dying with a hammer in his hand instead of losing to a steam drill~\cite{Henry} and to a psychologist Abraham Maslow stating that if the only tool you have is a hammer, it is tempting to treat everything as if it were a nail~\cite{Maslow}.}}}

\author{Vadim Zaytsev, \href{mailto:vadim@grammarware.net}{\texttt{vadim@grammarware.net}}\\[1em]
		\large
		Software Analysis and Transformation (SWAT) Team \\
		Centrum Wiskunde \& Informatica (CWI)\\
		Amsterdam, The Netherlands
		}

\usepackage[unicode,bookmarks=false,pdfstartview={FitH},%
            colorlinks,linkcolor=blue,urlcolor=blue,citecolor=blue,%
            pdfauthor={Dr. Vadim Zaytsev},%backref=page,%
            pdftitle={The Grammar Hammer of 2012}]{hyperref}
\begin{document}

\maketitle

%-------------------------------------------------------------------------

{\footnotesize\tableofcontents}

\input{intro}
\input{core}
\input{minor}
\input{venues}
\input{conclusion}

{
\newpage
\printbibliography[notkeyword=talk,notkeyword=unpublished,title={References}]
\newpage
\printbibliography[keyword=unpublished,title={Unpublished work}]
\printbibliography[keyword=talk,title={Presentations}]
}
\end{document}

%% file: intro.tex
\newpage
\section{Introduction}

% ===========================================================================
% \subsection{True intro}

% http://dx.doi.org/10.1038%2F35074210 !!! \cite{SelfArchiving}

The purpose of this report is documenting personal research results of the year 2012 in a form
primarily intended for assessment of their scientific merit as a foundation for future work, not for
quantitative assessment of the resulting publication record. This can be considered as an aggressive
form of self-archiving initiative~\cite{SelfArchiving} where scientific and engineering
contributions are not only logged, but also put in perspective by a separate first class atomic
scientific knowledge object. This report is mostly meant for my SWAT colleagues. However, it is open
for broad audience and meant to be readable by any researcher with reasonable degree of familiarity
with computer science. It can be consumed as a self-contained document, but many details are not
pulled in from available referenced sources.

We start right away with a the overview of the field (\S\ref{S:bg}) followed by brief descriptions
of major (\S\ref{S:major}) and minor (\S\ref{S:minor}) contributions, followed by a more elaborate
motivation for creation of this document (\S\ref{S:why}). Next, all research topics are laid out in
detail one by one (\S\ref{S:topics}). For the sake of complexity, a separate overview of all
involved venues (\S\ref{S:venues}) is included. \S\ref{S:end} concludes the report.

% ===========================================================================
\section{Preliminaries}
\subsection{Background notions}\label{S:bg}

\emph{Software language} is a concept that generalises over programming languages, markup languages,
database schemata, data structures, abstract data types, data types, modelling languages,
ontologies, etc. Whenever we observe some degree of \emph{commitment to structure}, we can identify
it with a language, which elements (symbols) can be separately defined and the allowed combinations
of them can be somehow specified. Studying software language engineering is important because of
possibly gained insights into relations between the way such languages are defined and used in
different technological spaces (e.g., we can study data binding as a way to map a relational
database to an object model, or language convergence as a way to compare an XML schema with a syntax
definition).

\newpage
\emph{Formal grammars} is a long-existing approach of dealing with languages --- originally context free
grammars~\cite{Chomsky56} were mainly aimed at textual programming languages~\cite{DragonBook}, but later
other variants of grammars were proposed, including
		keyword grammars~\cite{Meertens77},
		indexed grammars~\cite{Aho:1968:IGE:321479.321488},
		lexicalised grammars~\cite{Schabes:1988:PSL:991719.991757},
		object grammars~\cite{ObjectGrammars2012},
		pattern grammars~\cite{grenander1996elements},
		array grammars~\cite{SSK72},
		puzzle grammars~\cite{NSSSD91},
		% matrix grammars,
		picture grammars~\cite{MS67}, % picture calculus =?= picture grammars
		picture processing grammars~\cite{Chang:1970:ATP:800161.805166},
		tile grammars~\cite{Reghizzi:2005:TRG:1103398.1103405},
		grid grammars~\cite{YuPaun01},
		motion picture grammars~\cite{springerlink:10.1007/3-540-63931-4_228},
		pair grammars~\cite{Pratt:1971:PGG:1739929.1740026},
		triple graph grammars~\cite{TGG94},
		deterministic graph grammars~\cite{Caucal07},
		string adjunct grammars~\cite{JKY69},
		head grammars~\cite{Pollard84},
		tree adjunct grammars~\cite{JLT75},
		tree description grammars~\cite{springerlink:10.1023/A:1011431526022},
		description tree grammars~\cite{Rambow:1995:DG:981658.981679},
		description tree substitution grammars~\cite{Rambow:2001:DSG:972778.972782},
		functional grammars~\cite{DBLP:journals/ipl/Lukaszewicz77},
		{\L}ukaszewicz universal grammars~\cite{Lukaszewicz198276},
		two level grammars~\cite{Wijngaarden74},
		van Wijngaarden grammars~\cite{Wijngaarden65},
		metamorphosis grammars~\cite{springerlink:10.1007/BFb0031371},
		affix grammars~\cite{Koster91}, % reference to overview of all affix/attribute grammars!
		extended affix grammars~\cite{Meijer90},
		attribute grammars~\cite{AG-Genesis},
		extended attribute grammars~\cite{WM83},
		definite clause grammars~\cite{PW86},
		minimalist grammars~\cite{Lecomte:2001:ELG:1073012.1073059},
		categorial grammars~\cite{Ajdukiewicz35}, % Ajdukiewicz proposed; Lambek gave a calculus
		type grammars~\cite{Lambek58},
		pregroup grammars~\cite{Lambek08},
		Montague universal grammars~\cite{Montague70},
		logic grammars~\cite{DBLP:books/daglib/0067304},
		assumption grammars~\cite{Dahl97assumptiongrammars},
		constraint handling grammars~\cite{Christiansen05},
		abductive logic grammars~\cite{springerlink:10.1007/978-3-642-02261-6_14},
		simple transduction grammars~\cite{Lewis:1968:ST:321466.321477},
		inversion transduction grammars~\cite{Wu:1997:SIT:972705.972707},
		range concatenation grammars~\cite{boullier:inria-00073347},
		island grammars~\cite{DocGen},
		bridge grammars~\cite{Nilsson-Nyman:2009:PSR:1532448.1532458},
		skeleton grammars~\cite{Tolerant},
		permissive grammars~\cite{KJNV09},
		conjunctive grammars~\cite{Okhotin01},
		Boolean grammars~\cite{Okhotin200419},
		Peirce grammars~\cite{springerlink:10.1023/A:1011403527615}, % CFG + Pierce algebra as semantics
		transformational grammars~\cite{springerlink:10.1007/3540069585_50},
		probabilistic grammars~\cite{springerlink:10.1007/s10849-011-9135-z},
		notional grammars~\cite{Anderson91},
		analytic grammars~\cite{PEG},
		parsing schemata~\cite{DBLP:books/daglib/0085473},
		cooperating string grammar systems~\cite{CVDKP94},
		cooperating array grammar systems~\cite{DFP95},
		cooperating puzzle grammar systems~\cite{SRC06},
etc\footnote{The earliest possible reference is given for each variant, preferably from the programming language
research field.}. A grammar of a software language, which specifies commitment to grammatical structure, is called a
\emph{grammar in a broad sense}~\cite{KlintLV05}, even if in practice it defines a metamodel or an API, thus not
officially being a grammar at all. The grammarware technological space is commonly perceived as mature and drained of
any scientific challenge, but provides many unsolved problems for researchers who are active in that field.
% see also: http://en.wikipedia.org/wiki/Template:Formal_languages_and_grammars
% WTF COOL!!?!?!? http://en.wikipedia.org/wiki/Controlled_grammar
% http://en.wikipedia.org/wiki/Functional_grammar

For the last years, and specifically in 2012, I have focused my efforts on using grammar-based
techniques in the broad field of software language engineering.

% ---------------------------------------------------------------------------
\subsection{Major contributions in a nutshell}\label{S:major}

This section contains brief descriptions of the contributions of 2012 and some statements about their usability and/or
importance. Sections that contain extended descriptions of the contributions with some level of technical detail, are
referenced in parenthesis.

\begin{description}
	\item[Guided grammar convergence] (\S\ref{S:guided}).\\
	Grammar convergence is a lightweight verification method for establishing and maintaining the correspondence
	between grammar knowledge ingrained in various software artifacts~\cite{Convergence2009}.
	The method entails programming grammar transformation steps with a general purpose grammar transformation operator
	suite. It was acknowledged in \cite[p.34]{CamachoThesis} as ``a product-line approach to provide
	[...] an organised software structure''. Yet, the method had some weak sides that inspired
	further investigation.
	
	One of the biggest issues is maintenance of the grammar relationships. Once they have been
	established by programming grammar transformation steps, it becomes very hard to coevolve
	these steps with eventual changes in the source grammars. An ideal solution would be a way to
	automatically reestablish grammar relationships based on declarative constraints. This way is
	\emph{guided grammar convergence}: instead of programming the transformations, we construct an
	idealised ``master grammar'' that shows the most essential properties of all grammars that are
	to be converged, and the transformation steps are then derived automatically,
	guided by the structure of the master grammar.
	
	The transformation inference algorithm relies on the source grammars and their metasyntax. This
	method was prototyped twice: in Python and in Rascal, and tested successfully on 12 grammars in
	a broad sense obtained from different technological spaces. It has not been properly published
	after being rejected three times~\cite{Guided-ECMFA2012,Guided-ICSM2012,Guided-POPL2013},
	but received encouraging feedback from some of those venues and from one
	presentation~\cite{Guided-Convergence2012-talk}.
	\item[Grammar transformation] (\S\ref{S:trafo})\\
	Grammar convergence, evolution, maintenance and any other activity that deals with changes, can
	profit from expressing such changes in the functional way: every step is represented as a
	function application, where a function is a transformation operator such as \emph{rename} or
	\emph{add}. The latest of such operator suites has been developed in
	2010~\cite[XBGF Manual]{SLPS} and shown to be superior to its
	alternatives~\cite[\S4]{JLS-SQJ2011}.
	
	During 2012, XBGF has been:
	\begin{denselist}
		\item reimplemented in Rascal, which led to extensive testing and
		more systematic specification of operator semantics (\S\ref{S:XBGF});
		\item extended for
		bidirectionality by pairing operators, introducing lacking ones and abandoning unsuitable
		ones (\S\ref{S:CBGF});
		\item experimentally extended for adaptability (\S\ref{S:NBGF});
		\item extended by mining patterns of it usage (\S\ref{S:EXBGF});
		\item investigated for migration from the functional paradigm to the declarative one
		(\S\ref{S:3BGF}).
	\end{denselist}
	Each of these initiatives is a nontrivial project complete with conceptual motivation,
	programmed prototypes and obtained results (positive for the first three, controversial for the
	fourth and decisively negative for the last one).
	\item[Metasyntactic experiments] (\S\ref{S:meta})\\
	Metasyntax as a language in which grammars are specified, was a topic briefly touched in my
	PhD thesis~\cite{Zaytsev-Thesis2010}, but never officially published. In 2012, I finally
	dedicated enough time and attention to engineer a proper prototype for metasyntax
	specifications (\S\ref{S:EDD}) and their transformations (\S\ref{S:XEDD}), as well as to perform
	a series of experiments on metasyntax-driven grammar recovery (\S\ref{S:EDDrec}) and
	convergence (\S\ref{S:EDDguided}). This area has now been exhaustively covered,
	and the only possible future extensions must rely on going way beyond textually specified
	context-free grammars.
	
	To be completely frank, it should be noted here that most of the experiments with metasyntax
	were done in the course of 2011 and were only polished, presented and published in 2012 (which still
	required considerable effort).
	\item[Tolerant parsing overview] (\S\ref{S:tolerant})\\
	Just like the grammar recovery paper came with an extensive related work section which listed
	all grammar recovery initiatives in the last decade or two~\cite[\S2]{NPGR2012}, a new parsing
	algorithm that I tried to propose (\S\ref{S:parsing}) came with an extensive overview of all
	methods of tolerant parsing known to grammarware engineers up to date (\S\ref{S:tolerant}).
	While the iterative parsing method was novel but ultimately dull and uninteresting, the overview
	itself was received very warmly during the presentation on it~\cite{Tolerance2012-talk}.
	One of the reviewers of \cite{Islands-SCAM2012} has also advised to throw away the thing I
	thought was the main contribution of the paper, and extend the thing I thought of as a
	byproduct, into a longer journal article. While surprising at first, this seems indeed like a
	reasonable course of action.
\end{description}

% ---------------------------------------------------------------------------
\subsection{Selected minor contributions}\label{S:minor}

\begin{figure*} % htbp
	\centering
		\includegraphics[width=.8\textwidth]{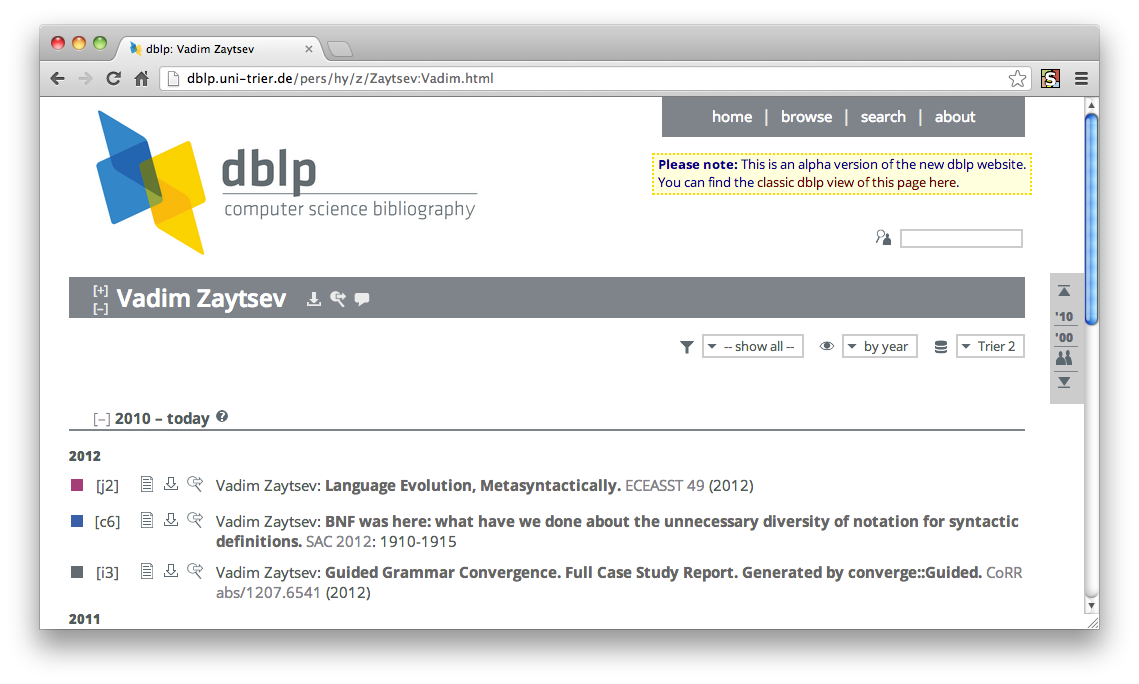}
	\caption{The results of 2012, according to DBLP.}
	\label{F:dblp}
\end{figure*}

In the following sections, I will present a detailed overview of major (\S\S\ref{S:guided}--\ref{S:ONS})
and minor (\S\ref{S:minortopics}) contributions, but the border between them is naturally flexible. Thus,
in the previous section introduced only four of the best major ones, and this section will introduce
several middleweight contributions (``less major'' mixed with ``not so minor'' ones).

\begin{description}
	\item[Grammar mutation] (\S\ref{S:mutation})\\
	It has been noted in \cite{Metasyntactically-BX2012,Metasyntactically2012} that there is a separate
	group of grammar changes that reside between traditional grammar transformations (``rename X to Y'')
	and the grammar transformation operators (``rename''), which was labelled as a grammar \emph{mutation}
	and formalised differently from them. While the only truly important property of grammar mutation in
	the context of \cite{Metasyntactically-BX2012,Metasyntactically2012} was that they are considerably
	harder to bidirectionalise, a lot of useful grammar manipulations like ``rename all uppercase
	nonterminals to lowercase'' or ``eliminate all nonterminals unreachable from the root'' belong to the
	class of mutations, so it deserves to be studied closer. In \cite{Trends2012}, I have composed a list
	of 16 mutations identified in already published academic papers or in publicly available grammarware
	source code, but the paper was not accepted, so the topic remains only marginally explored.
	\item[Iterative parsing] (\S\ref{S:parsing})\\
	Starting from a fresh yet weird topic of what ``the cloud'' can mean for grammarware engineering, I
	ended up proposing an algorithm for \emph{parsing in the cloud}, which was \textbf{not} based on
	parallel parsing~\cite{Alblas:1994:BPP:181577.181586}, but rather on island
	grammars~\cite{Moonen01,Tolerant}. The whole topic is questionable and only suitable for a ``wild
	ideas workshop'', as was nicely put by one of the reviewers, but is still potentially of some
	interest. The paper containing the algorithm was rejected
	twice~\cite{Islands-SCAM2012,Islands-NordiCloud2012} so far, and requires investing more time in
	empirical validation at least, in order to increase the chances of acceptance.
	\item[Unparsing in a broad sense] (\S\ref{S:unparsing})\\
	I could not help noticing that parsing (i.e., mapping strings to graphs) receives much more research
	attention than the reverse process of unparsing (i.e., mapping graphs to strings). However, the only
	thing I did accomplish this year was to collect a couple of references on existing research and
	make a ``new ideas'' extended abstract~\cite{Unparsing-Techniques2012}, which was classified as a
	``request for discussion'' and rejected. I am already prepared to give a discussion-provoking
	presentation on this topic, but it requires much more effort to be invested until more tangible
	results are obtained.
	\item[Megamodelling] (\S\ref{S:mega})\\
	Megamodelling is higher abstraction level form of modelling that is concerned with software languages
	and technologies and relations between them. This year I have published some papers with megamodels in
	them~\cite{Negotiated-XM2012,Negotiated2012,Renarration-MPM2012,Renarration2012,Guided2012} and
	touched upon the topic in a range of
	presentations~\cite{MegaL2012-talk,Renarration-SLT2012-talk,Renarration-MPM2012-talk,Negotiated-XM2012-talk}. Much more work on this topic is planned for 2013.
	\item[Open notebook computer science] (\S\ref{S:ONS})\\
	Open notebook science is an open science paradigm of doing research in a transparent way.
	It is already a fairly widely accepted methodology in areas like chemistry~\cite{DoD2008} and drug
	discovery~\cite{Singh2008201} and is generally perceived as the next big step after open
	access~\cite{Lloyd2008}. However, in computer science and software engineering it has never been
	a tradition to keep a lab notebook, and it takes quite some time to maintain it, with few apparently
	visible benefits. I have been experimenting quite a lot with this idea, but finally decided to come
	out to a bigger public with two presentations in 2012~\cite{Open2012-talk,Subatomic2012-talk}. In
	general, I believe this is a reasonable idea, and I will keep practicing open notebook science myself,
	but it will take quite some effort to put it carefully into words in order to publish, so I am not
	even sure it is feasible to expect a publication in 2013.
	% \item[Structured data extraction].\\
	% ...
\end{description}

% ===========================================================================
\subsection{Motivation for this report}\label{S:why}

\begin{figure*} % htbp
	\centering
		\includegraphics[width=.7\textwidth]{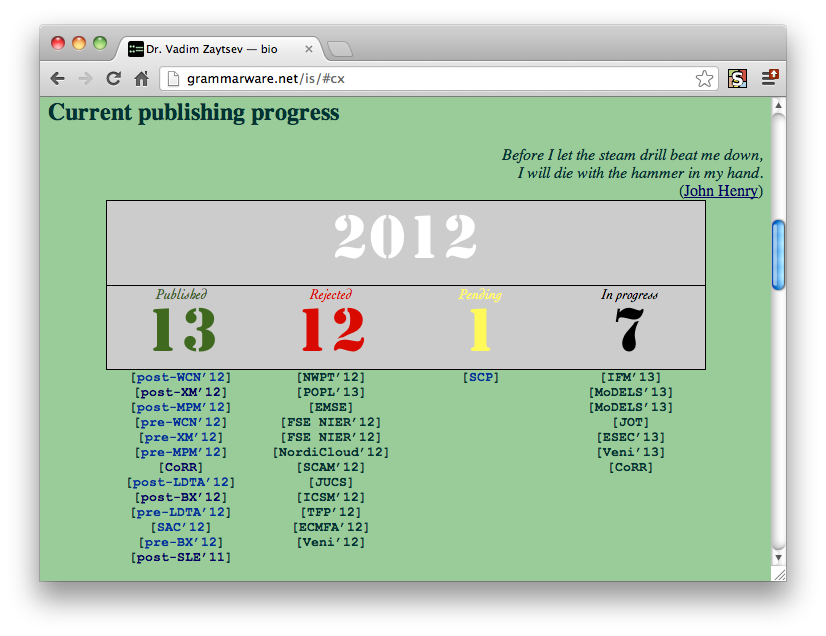}
	\caption{The results of 2012, according to the self-archiver.}
	\label{F:final}
\end{figure*}

The progress of a scientist is traditionally measured by an outsider by the papers that the scientist
produces. According to DBLP, the main supplier of bibliography lists currently, the year 2012 for me
yielded the following results (see the screencapture on \autoref{F:dblp}): one journal
paper~\cite{Metasyntactically2012}, one conference proceedings paper~\cite{BNF-WAS-HERE2012}, one
preprint~\cite{Guided2012}. However, the first one is an only slightly extended version of a workshop
paper~\cite{Metasyntactically-BX2012} written mostly in 2011; the second one was written and accepted in
2011; and the third one was intended to be a supplementary material for another paper that is not yet
accepted anywhere. Additionally, there are three more post-proceedings papers in
print~\cite{NPGR2012,Renarration2012,Negotiated2012}, which are already finished and submitted and will
eventually appear in the ACM Digital Library --- when they do, they will also be listed at DBLP under
2012, but at that time it will be too late to write a year report.

What about the self-archiving initiative~\cite{SelfArchiving}? Luckily, I disclose relatively large
amounts of dark data~\cite{DarkData2007} about my research activities, having an extensively linked daily
updated website with an open notebook (see \S\ref{S:ONS}) and many generated lists, including the current
publishing progress, as seen on the the screencapture on \autoref{F:final}. Even judging by the bare
numbers, one can already tell that this list contains much more information than the DBLP list. However,
it also has its problems: the ``published'' column contains the works of previous years that happened to
be delayed enough for the post-proceedings to appear in January 2012~\cite{TestMatch2012}; as well as
mentions of drafts planned for future publication (easily localised in the last column). It also contains
editorial work for non-mainstream venues~\cite{WCN2012,WCNe2012} which is of much lesser relevance because
there is no scientific value to it. What it does not contain, is relations between all these papers:
obviously some papers are enhanced version of previously rejected drafts, but in order to figure them out,
one needs to read the open notebook at \url{http://grammarware.net/opens} or analyse it automatically (no
readily available tools are provided).

Personally, I can state that guided grammar convergence (see \S\ref{S:guided}) is my top result of the
year. However, it has not (yet) been properly published. After being rejected at
ECMFA~\cite{Guided-ECMFA2012} and ICSM~\cite{Guided-ICSM2012}, it received very positive reactions from
POPL~\cite{Guided-POPL2013}, yet was also deemed not mature enough for publication. Still, having to
figure out what are the limits of the proposed methodology and how to describe it well, does not change
the fact that this is my best contribution of the year 2012.

Grammar transformation operator suites like XBGF (see \S\ref{S:XBGF}), \CBGF\ (\S\ref{S:CBGF}), EXBGF
(\S\ref{S:EXBGF}), $\Delta$BGF (\S\ref{S:3BGF}) and NBGF (\S\ref{S:NBGF}) represent massive amounts of
work, but they are not publishable by themselves, if at all. Still, each of them represents a milestone
enabling further advances. Engineering work that supports scientific research, has rarely been explicitly
noted and appreciated.

Quoting \cite{DoD2008}: \emph{``The notebook is about publishing data as quickly as possible. The paper is
about synthesizing knowledge from all those results.''} Hence, this report is aimed at synthesizing
knowledge about the experiments and achievements undertaken during the course of 2012 by me (possibly in
collaboration with someone else) within the NWO project 612.001.007, ``Foundations for a Grammar
Laboratory''. It holds the most value for myself and my project colleagues, but is also available for
anyone interested in the topics discussed: unlike open notebook entries, this report is a proper atomic
scientific knowledge object~\cite{Giunchiglia:2010:SKO:2328909.2328928,Liquid2011}. Only two topics
directly relevant to the project, are not included: one must remain hidden according to the rules of the
target venue, and for the other one the context and consequences are not yet understood enough even for
such a lightweight presentation.

%% file: core.tex
\section{Topics overview}\label{S:topics}

% ===========================================================================
\subsection{Guided grammar convergence}\label{S:guided}

Let us consider two grammars in a broad sense \cite{KlintLV05}. We say that they represent one
\emph{intended} software language, if there exists a complete bidirectional mapping between language
instances that commits to grammatical structure of different grammars. For example, if a parser produces
parse trees that can always be converted to abstract syntax trees expected by a static analysis tool and
back, it means that they represent the same intended language. As another example, consider an object
model used in a tool that stores its objects in an external database (XML or relational): the existence of
a bidirectional mapping between entries (trees or tables) in the database and the objects in memory, means
that they represent the same intended language, even though they use very different ways to describe it.
An equivalence class spawned by this definition (i.e., a set of different grammars of the same intended
language) effectively forms a \emph{grammarware product line} of products that perform different tasks
on instances of the same intended language: in that sense, for example, all Java-based tools form a
product line, if they agree on a language version and do not employ any highly permissive methods that
would shift them into a broader class. For the sake of simplicity, let us focus on \emph{grammar product
lines}: collections of grammars of the same intended language. The relation between a grammar product line
and a grammarware product line is justified by research on automated derivation of grammar-based tools
like parsers, environments, documentation, formatters and renovators from
grammars~\cite{ASFSDF-Klint,DeYacc,PrettyPrint3,KlintLV05,CamachoJournal,LDF2011}.

Suppose that we have two grammars: one that we call a \emph{master grammar} (a specially pre-constructed
abstract grammar of the intended language) and one that we call a \emph{servant grammar} (a grammar
derived from a particular language implementation). In general, there are four phases of guided grammar
convergence, and they are presented in this section in the \emph{reverse} order. First, we consider the
simplest scenario when all mismatches are of \emph{structural} nature. Then, we move on to a more
complicated situation when a \emph{nominal} matching between sets of nonterminals is unknown. Since this
is rather uncommon (most methods used in practice for imploding parse trees to abstract syntax trees, from
Popart~\cite{Wile1997} to Rascal~\cite{Rascal}, heavily rely on equality of names), a new method for
matching nonterminals has been developed. In short, it comprises construction of production signatures for
each production rule in both grammars, and a search for equivalent and weakly equivalent production rules
with respect to those signatures. Once a name resolution relation has been successfully built, a
previously discussed structural matching can be applied. We will also discuss \emph{normalisations} that
can transform any arbitrary grammar to a form easily consumable by our nominal and structural matching
algorithms. Finally, I will list additional problems that indicate \emph{grammar design} decisions and
therefore not affected by normalisations. However, I describe how to automatically detect such issues and
to address them with grammar mutations.

\subsubsection*{Structural matching}

Let us assume the simplest scenarios: the two input grammars have the same set of nonterminals; neither of
them has terminals; the starting nonterminal is the same and that the sets of production rules are
different but have the same cardinality. These would be typical circumstances if, for example, the
grammars define two alternative abstract syntaxes for the same intended language.

We can start from the roots of both grammars and traverse them synchronously top-down, encountering only the following four circumstances:

\begin{description}
	\item[Perfect match.] Convergence is trivially achieved.
	\item[Nonterminal vs.\ value.] By ``values'' I mean nonterminals that are built-in in the underlying
	framework (e.g., ``string'').
	\item[Sequence element permutations] can be automatically detected and converged.
	\item[Lists of symbols.]  Many frameworks that have components with grammatical knowledge, have a
	notion of a list or a repetition of symbols in their metalanguage.
\end{description}

It can be shown that these four are the only possibilities, and that their resolution can be resolved.

\subsubsection*{Nominal resolution}

In a more complicated scenario, let us consider the case of different nonterminal sets in two input
grammars, and for simplicity we assume that all production rules are vertical (non-flat) and chained (if
there is more than one production rule for the same nonterminal, all of them are chain productions ---
i.e., have one nonterminal as their right hand side). Next, we define a \emph{footprint} of a nonterminal
in an expression as follows:

$$ \pi_n(x) =
\begin{cases}
	\{1\}	& \text{if } x=n\\
	\{?\}	& \text{if } x=n?\\
	\{+\}	& \text{if } x=n^{+}\\
	\{*\}	& \text{if } x=n^{*}\\
	\bigcup\limits_{e\in L} \pi_n(e)	& \text{if } x\text{ is a sequence }L\\
	\varnothing	& \text{otherwise}
\end{cases} $$

By extension, we define a footprint of a nonterminal in a production rule as a footprint of it in its right hand side:

$$ \pi_n(m \to e) = \pi_n(e) $$

Based on that, we define a \emph{production signature}, or a prodsig, of a production rule, by collecting all footprints of all nonterminals encountered in its right hand side:

$$\sigma(p) = \{\langle n, \pi_n(e) \rangle\: |\: n\in\mathbb{N},\:\pi_n(e) \not= \varnothing \}$$

\begin{table}\footnotesize
\centerline{\begin{tabular}{|l|c|}\hline
	\multicolumn{1}{|>{\columncolor[gray]{.9}}c|}{\footnotesize \textbf{Production rule}} &
	\multicolumn{1}{>{\columncolor[gray]{.9}}c|}{\footnotesize \textbf{Production signature}}
	\\\hline
	$p_1$=(\emph{program} $\to$ \emph{function}$^{+}$) & $\{\langle \textit{function}, \plus \rangle\}$ \\
	$p_2$=(\emph{function} $\to$ $str$ $str^{+}$ \emph{expr}) & $\{\langle \textit{expr}, {1} \rangle, \langle str, {1}\plus \rangle\}$ \\
	$p_3$=(\emph{expr} $\to$ $str$) & $\{\langle str, {1} \rangle\}$ \\
	$p_4$=(\emph{expr} $\to$ $int$) & $\{\langle int, {1} \rangle\}$ \\
	$p_5$=(\emph{expr} $\to$ \emph{apply}) & $\{\langle \mathit{apply}, {1} \rangle\}$ \\
	$p_6$=(\emph{expr} $\to$ \emph{binary}) & $\{\langle \mathit{binary}, {1} \rangle\}$ \\
	$p_7$=(\emph{expr} $\to$ \emph{cond}) & $\{\langle \mathit{cond}, {1} \rangle\}$ \\
	$p_8$=(\emph{apply} $\to$ $str$ \emph{expr}$^{+}$) & $\{\langle \textit{expr}, \plus \rangle, \langle str, {1} \rangle\}$ \\
	$p_9$=(\emph{binary} $\to$ \emph{expr} \emph{operator} \emph{expr}) & $\{\langle \textit{expr}, {1}{1} \rangle, \langle \textit{operator}, {1} \rangle\}$ \\
	$p_{10}$=(\emph{cond} $\to$ \emph{expr} \emph{expr} \emph{expr}) & $\{\langle \textit{expr}, {1}{1}{1} \rangle\}$ \\
	\hline
\end{tabular}}
\caption{Production rules of the master grammar for FL, with their production signatures.}
\label{F:prodsigs}
\end{table}

A good example of how production signatures look like, is to be found on \autoref{F:prodsigs}.

We say that two production rules are \emph{prodsig-equivalent}, if and only if there is a unique match
between tuple ranges of their signatures:
\vspace{-.5em}
$$p \bumpeq q \: \Longleftrightarrow \:
	\forall\langle n,\pi\rangle\in\sigma(p),\:
	\exists! \langle m,\xi\rangle\in\sigma(q),\:
	\pi=\xi$$

Similarly, a weak prodsig-equivalence $p \Bumpeq q$ is defined by dropping the uniqueness constraint and
weakening the equality constraint in the last definition to footprint equivalence which disregards
repetition kinds ($\plus$ is equivalent to $\star$). Then it can be proven that for any two strongly
prodsig-equivalent production rules $p$ and $q$, $p\bumpeq q$, a \emph{nominal resolution} relationship
has the form of:

$$p \diamond q = \sigma(p) \circ \overline{\sigma(q)} $$

where $\rho_1 \circ \rho_2$ is a composition of two relations in the classic sense and $\overline{\rho}$
is the classic inverse of a relation. Moreover, for any two weakly prodsig-equivalent production rules $p$
and $q$, $p\Bumpeq q$, there is (at least one) nominal resolution relationship $p \diamond q$ that
satisfies the following:
\vspace{-1em}
\begin{gather*}
\forall \langle a,b\rangle \in p \diamond q:
a = \omega \vee b = \omega \:\vee\\
\exists \pi, \exists \xi,
\pi\approx\xi,
\langle a,\pi\rangle\in\sigma(p),
\langle b,\xi\rangle\in\sigma(q)
\end{gather*}

and \vspace{-1em}

$$
\forall \langle a,b\rangle \in p \diamond q,
\forall \langle c,d\rangle \in p \diamond q:
a = c \Rightarrow b = d
$$

Where $\omega$ is used to explicitly denote unmatched nonterminals.

\subsubsection*{Abstract Normal Form}

In order to fit any grammar into the conditions required by the previously described matching techniques,
we demand the following normalisation:

\begin{enumerate}
	\item lack of labels for production rules
	\item lack of named subexpressions
	\item lack of terminal symbols
	\item maximal outward factoring of inner choices
	\item lack of horizontal production rules %nonterminals (those which definition is a top level choice)
	\item lack of separator lists
	% separator lists are desugared to simple repetitions
	\item lack of trivially defined nonterminals (with $\alpha$, $\varepsilon$ or $\varphi$)
	\item no mixing of chain and non-chain production rules
	% \item lack of chain production rules (of a form `$a \to b$', where $a\in\mathbb{N}, b\in\mathbb{N}$)
	% \item the set of starting symbols equals the set of top nonterminals connected to the rest of the grammar
	\item the nonterminal call graph is connected, and its top nonterminals are the starting symbols of the grammar
\end{enumerate}

It can be shown that transforming any grammar into its Abstract Normal Form is in fact a grammar mutation
(see \S\ref{S:mutation}). In the prototype, I have implemented it to effectively generate bidirectional
grammar transformation steps, so the normalisation preserves any information that it needs to abstract
from.

\subsubsection*{Grammar design mutation}

Some grammar design smells (terminology per \cite{PEM7}) like yaccification (per \cite{DeYacc,Harmful}) or
layered expressions (per ~\cite{Convergence2009}) have shown to be persistent enough to survive all
normalisations and cause problems for establishing nominal and structural mappings. They can be identified
and dealt with by automated analyses and mutations, but so far I have to proof that they are the only
possible obstacles, and no guarantees about any other smells problematic for guided grammar convergence.

\subsubsection{Generalisation of production signatures}

The method of establishing nonterminal mappings of different grammars of the same intended language, can
be generalised as follows. Suppose that we have a metalanguage. Without loss of generality, let us assume
that each grammar definition construct that is present in it, can be referred to by a single symbol:
``\ccomma'', ``\textbf{?}'', ``\textbf{*}'', etc and uses prefix notation. This metasyntactic alphabet
$\Lambda$ will form the foundation of our footprints and signatures. Let us also assume that all
metasymbols are unary or are encoded as unary, except for two composition constructs: a sequential
``\ccomma'' and an alternative ``$\mathbf{|}$'', which take a list of symbols.

Then, a footprint of any nonterminal $n$ in an expression $x$ is a multiset of metasymbols that are used
for occurrences of $n$ within $x$:

 $$ \pi_n(x) = \begin{cases} \{1\} & \text{if } x=n\\ \{\mu\} & \text{if } x=\mu(n), \mu\in\Lambda\\
\bigcup\limits_{e\in L} \pi_n(e) & \text{if } x=\ccomma(L)\\ \varnothing & \text{otherwise, also if }
x=\mathbf{|}(L)\\ \end{cases} $$

Our previously given definition of a production signature can still be used with this generally redefined
footprints.

It is well known that language equivalence is undecidable. Any formulation of the grammar equivalence
problem, that is based on language equivalence, is thus also undecidable. Grammar
convergence~\cite{Convergence2009,JLS-SQJ2011} is a practically reformulated grammar equivalence problem
that uses automated grammar transformation steps programmed by a human expert. By using these generalised
metasyntactic signatures, we can \emph{infer converging transformation steps} automatically, thus
eliminating the weakest link of the present methodology. However, this is not the only application of the
generalisation.

The most trivial use of metasyntactic footprints and signatures would lie in \emph{grammarware metrics}.
Research on software metrics applied to context-free grammars has never been an extremely popular topic,
but it did receive some attention in the 1970s~\cite{Gruska}, 1980s~\cite{Kelemenova1981} and even
recently~\cite{PowerMalloy,CrepinsekMernik}. Using quantitative aspects of metasyntactic footprints and
signatures (numbers of different footprints within the grammar, statistics on them, etc) is possible and
conceptually akin to using micropatterns~\cite{Gil:2005:MPJ:1094811.1094819} and
nanopatterns~\cite{Batarseh:2010:JNP:1900008.1900089}, but nothing of this kind has ever been done for
grammars (in a broad sense or otherwise).

A different more advanced application of metasyntactic footprints and signatures is the analysis of their
usage by mining existing grammar repositories like Grammar Zoo~\cite{SLPS}. This can lead to not only
improving the quality of the grammars by increasing their utilisation of the metalanguage functionality,
but also to \emph{validation of metalanguage design}. The whole programming language community uses
dialects and variations of BNF~\cite{BNF} and EBNF~\cite{EBNF}, but their design has never been formally
verified. However, one may expect that introducing EBNF elements like symbol repetition to BNF can be
justified by analysing plain BNF grammars and finding many occurrences of encoding them
(``yaccification'', etc). It will also be interesting to see what new features the EBNF lacks practically
--- none of the existing proposals so far (ABNF~\cite{ABNF}, TBNF~\cite{TBNF}, etc) were ever formally
validated.

\subsubsection{History of attempted publication}

Initially, the idea of guided grammar convergence has emerged as a contribution for
ECMFA~\cite{Guided-ECMFA2012}. The level of contribution was praised by the reviewers, but the paper
itself was deemed inappropriate for a heavily model-related venue. A bit later it was resubmitted after
minor revision to ICSM~\cite{Guided-ICSM2012}, where it was received even colder, presumably because the
reviewers were seeking a more practical side which was not demonstrated well enough. After much more
effort put into experiments, prototypes, auxiliary material~\cite{Guided2012} and a complete rewrite of
the paper itself, the method was submitted to POPL~\cite{Guided-POPL2013}. It was unanimously rejected,
but with very constructive and encouraging reviews. In 2013, they will be taken into account when the
paper will be submitted again (the last time as a conference paper --- otherwise I will admit it to be
impossible for me to explain this method within the common limitations and go for a much longer
self-contained journal submission).

In \cite{Incremental2012}, I have attempted to sell the very act of validating the new method of guided
grammar convergence by letting it cover the older case study done with contemporary grammar convergence,
as a some sort of experimental replication in a broad sense. The reviewers praised the nonconformism and
originality of the approach, and rejected the paper.

The generalisation of the method was proposed as an extended abstract to NWPT~\cite{Metasyntactic2012},
where the reviewers did not see any merit in it (which I personally found strange since both ICSM and POPL
reviewers insisted that various components of the method like ANF and prodsigs must be treasured as
standalone contributions which applicability is much wider than the automated convergence of grammars).
Either my way of explaining was bad enough to obfuscate this point, or I have terribly misunderstood their
call for papers.

% ===========================================================================
\subsection{Grammar transformation languages}\label{S:trafo}

% ---------------------------------------------------------------------------
\subsubsection{XBGF}\label{S:XBGF}

XBGF, standing for Transformation of BNF-like Grammar Format, is a domain-specific language for
automated programmable operator-based transformations of grammars in a broad sense. It has been
previously implemented in Prolog (which was mostly done by Ralf L{\"a}mmel) and published as a part
of a journal article~\cite[\S4]{JLS-SQJ2011}, as well as a separate online manual~\cite[XBGF
Manual]{SLPS} --- in fact, just a byproduct of the research on language
documentation~\cite{LDF2011}.

XBGF is essentially finished work: it is working, it is useful for experiments, it has
documentation, it has a test suite, etc. The only thing that was added in the course of 2012 is the
reimplementation of XBGF in Rascal~\cite{Rascal}. Beside some metaprogramming, this reimplementation
led to streamlining some of the applicability preconditions and postcondition, which could be viewed
as a very minor scientific contribution.

% ---------------------------------------------------------------------------
\subsubsection{\CBGF}\label{S:CBGF}

If XBGF was read as ``iks bee gee eff'', then \CBGF\ is ``ksee bee gee eff'', its bidirectional
counterpart. Inspired by the call for papers of BX'12 (The First Workshop on Bidirectional
Transformations, see \S\ref{S:yesvenues}), I was experimenting with bidirectionality in the
grammarware technological space, and this language is what came out of it. 80\% of the work for
creating it involved trivial coupling of grammar transformation operators like \emph{chain} and
\emph{unchain}, but the remaining 20\% have provided a lot of fuel for thinking about what seemed to
be a polished and finished product. \CBGF\ was published as a part of online pre-proceedings
\cite{Metasyntactically-BX2012}, and then, after the second round of reviews, as a journal
article~\cite{Metasyntactically2012}. The only problem was that the BX paper took off on its own, so
the bidirectional grammar transformation operator suite seems like one of many byproducts there.
There was a failed attempt to craft a paper that would be more focused on \CBGF\ (and other aspects
of grammar transformation not covered sufficiently by the BX submission), but a wrong venue was
targeted, which resulted in desk rejection~\cite{Trends2012}.

% ---------------------------------------------------------------------------
\subsubsection{NXBGF?}\label{S:NBGF}

Another property of programmable grammar transformations that always bothered me, was their
rigidity: once written, they are hard to maintain and adapt, and one little change in the original
grammar (for example, when the extractor is changed) can unexpectedly and unpredictably break (make
defunct) some of the transformation steps much later in the chain, and there is no method available
to detect the change impact. Analysing this problem led to an idea that was originally in
preparation for the FM+AM workshop (see \S\ref{S:novenues}), but was not ready before the deadline,
so it went to the Extreme Modelling Workshop instead, where it received surprisingly warm reaction.

The idea is: negotiations. Whenever an error arises (usually an applicability condition is not met),
instead of failing the whole chain, try to recover by negotiating the outcome with the data about
near-failure and some external entity (usually an oracle or a human operator). For example, when we
want to rename a nonterminal that does not exist, the transformation engine may seek nonterminals
with names similar to the required one, and try renaming them.

The idea of negotiated grammar transformations was published in the online
proceedings~\cite{Negotiated-XM2012} and then in the ACM Digital Library~\cite{Negotiated2012},
after which I was invited to submit an extended version to a journal. This will soon lead to a
prototype implementation of such a system and perhaps to some interesting experiments with it. If
this advancement yields a yet another grammar transformation operator suite, it may or may not be
named ``NXBGF''.

% ---------------------------------------------------------------------------
\subsubsection{EXBGF}\label{S:EXBGF}

\begin{table*}
	\begin{center}
		\begin{tabular}{|l|r|r|r|r|r|r|r|r|}
\hline
~            & \textbf{jls1} & \textbf{jls2} & \textbf{jls3} & \textbf{jls12} & \textbf{jls123} & \textbf{r12} & \textbf{r123} & \textbf{Total} \\ \hline
XBGF, LOC    & 682           & 6774          & 10721         & 5114           & 2847            & 1639         & 3082          & 30859          \\ 
EXBGF, LOC   & 399           & 5509          & 7524          & 3835           & 2532            & 1195         & 2750          & 23744          \\ 
~            & $-$42\%         & $-$19\%         & $-$30\%         & $-$25\%          & $-$11\%           & $-$27\%        & $-$11\%         & $-$23\%          \\ 
genXBGF, LOC & 516           & 5851          & 9317          & 4548           & 2596            & 1331         & 2667          & 26826          \\ 
~            & $-$24\%         & $-$14\%         & $-$13\%         & $-$11\%          & $-$9\%            & $-$19\%        & $-$13\%         & $-$13\%          \\
\hline
XBGF, nodes  & 309           & 3,433         & 5,478         & 2,699          & 1,540           & 786          & 1,606         & 15851          \\ 
EXBGF, nodes & 177           & 2,726         & 3,648         & 1,962          & 1,377           & 558          & 1,446         & 11894          \\ 
~            & $-$43\%         & $-$21\%         & $-$33\%         & $-$27\%          & $-$11\%           & $-$29\%        & $-$10\%         & $-$25\%          \\ 
genXBGF, nodes& 326          & 3,502         & 5,576         & 2,726          & 1,542           & 798          & 1,610         & 16080          \\
~            & +6\%          & +2\%         & +2\%         & +1\%          & +0.1\%           & +2\%        & +0.3\%         & +1\%          \\ 
\hline
XBGF, steps    & 67      & 387     & 544     & 290     & 111     & 77      & 135     & 1611    \\
EXBGF, steps   & 42      & 275     & 398     & 214     & 98      & 50      & 120     & 1197    \\ 
...pure EXBGF      & 27      & 104     & 162     & 80      & 30      & 34      & 44      & ~       \\ 
...just XBGF       & 15      & 171     & 236     & 134     & 68      & 16      & 76      & ~       \\ 
~                & $-$37\% & $-$29\% & $-$27\% & $-$26\% & $-$12\% & $-$35\% & $-$11\% & $-$26\% \\ 
genXBGF, steps & 73      & 390     & 555     & 296     & 112     & 83      & 139     & 1648    \\ 
~                & +9\%   & +1\%   & +2\%   & +2\%   & +1\%   & +8\%   & +2\%   & +2\%   \\
\hline

\end{tabular}
	\end{center}
	\caption{Size measurements of the Java grammar convergence case study, done in XBGF and in EXBGF.
	In the table, XBGF refers to the original transformation scripts, EXBGF
	to the transformations in Extended XBGF, genXBGF measures XBGF scripts generated from EXBGF.
	LOC means lines of code, calculated with \texttt{wc -l}; nodes represent the number of nodes in
	the XML tree, calculated by XPath; steps are nodes that correspond to transformation operators
	and not to their arguments. Percentages are calculated against the XBGF scripts of the original
	study.}
	\label{T:EXBGF}
\end{table*}

Considerations about the state of XBGF led me to start cursory reexamination of the available
transformation scripts. The Java case study undertaken in 2009--2010 and published as a conference
paper~\cite{JLS-SCAM2009}, a journal paper~\cite{JLS-SQJ2011} and open source
repository~\cite{SLPS}, provided me with plenty of them. Manual ad hoc pattern recognition has
resulted in development of a new operator suite, with higher order operators such as
\emph{exbgf:pull-out}, which would be equivalent to a superposition of \emph{xbgf:horizontal},
\emph{xbgf:factor}, \emph{xbgf:extract} and \emph{xbgf:vertical}. As shown on \autoref{T:EXBGF},
size metrics show a drop of {23--26\%} in Extended XBGF with respect to XBGF, but also the
complexity was obviously decreased. However, the results were not extremely convincing and lacked
real strength since only a few uses per high level operator were found, and the new EXBGF language
was not designed systematically. Besides all that, the case study I have done, is, strictly
speaking, about \emph{refactoring} XBGF scripts to Extended XBGF, so claims about usefulness of
EXBGF for \emph{creating} new transformation scripts, should be stated with caution.

EXBGF was first described as an idea as a part of \cite{Trends2012}. After its rejection, it was
developed further and laid out in much more detail in a journal submission, which was also
eventually rejected~\cite{Incremental2012}. The fact that I presented Extended XBGF first as a
``trend'' and then as an ``experiment'', perfectly reflects my point of view that it is not a solid
contribution on its own.

% ---------------------------------------------------------------------------
\subsubsection{$\Delta$BGF?}\label{S:3BGF}

If there was one good outcome of getting a grammar transformation paper~\cite{Trends2012} rejected
at a functional programming conference, then this is it: I started contemplating how to specify them
in a non-so-functional way. Having recently been to a bidirectional transformations workshop helped,
and I started researching \emph{tridirectional} transformations (in fact, they quickly turned
multidirectional). The idea was clean and simple: do not specify grammar changes as functions;
instead, specify them as predicates. Such a predicate would, for example, introduce a nominal
binding between nonterminals in different grammars --- after which, the actual renaming steps can be
easily inferred from such a binding predicate.

Unfortunately, this idea was so beautiful in theory, but proven nearly impossible in practice (or in
detailed theory, for that matter). The main problem lies with the order of execution: a functional
grammar transformation script specifies that order naturally, while a list of predicates does not.
As I found out the hard way, my prototypes were still clean and beautiful when they dealt with one
transformation step; reasonable tricks and extensions could let me go up to three steps; beyond that
some serious redesign was needed; and so far I have not figured out how to overcome this.

% ===========================================================================
\subsection{Metasyntax}\label{S:meta}

Whenever we have a software language, we can speak of its \emph{syntax} as a way it allows and
disallows structural combinations of elements: programming languages rely on keywords and possibly
layout conventions; spreadsheets have ways of distinguishing between cells and referring to one from
another; markup languages have symbol sequences of special meaning; musical notes are arranged on a
grid; graphs must have uniquely identifiable nodes and edges connecting exactly two each; etc. Then,
a \emph{metasyntax} is a way of specifying this syntax. In the classic programming language theory,
languages are textual and can be processed as sequences of lexems, and the metasyntax is Backus
Normal Form~\cite{BNF}, also called Backus Naur Form~\cite{BNFvsEBNF}, or its enhanced variant
Extended Backus Naur Form~\cite{EBNF}. Despite the fact that EBNF has been standardised by
ISO~\cite{ISO-EBNF}, there is no agreement in the software language engineering community on the
exact variant of EBNF: some people just prefer using ``$:\equiv$'' or ``$\triangleq$'' instead of
``='' for esthetic reasons or prefer separating production rules with double newlines for
readability reasons and for the sake of easy processing.

The idea was hinted in my PhD thesis in 2010~\cite{Zaytsev-Thesis2010}, completely worked out in 2011 and was
put to several good uses in 2012. These are listed in the following subsections.

% ---------------------------------------------------------------------------
\subsubsection{Notation specification}\label{S:EDD}

The first step in treating metalanguages as first class entities is, of course, encapsulating a particular
metalanguage with a specification that defines it. By extending the list of possible metasymbols from the
ISO EBNF standard~\cite{ISO-EBNF} and by reusing the empirically constructed Table 6.1 from my
thesis~\cite[p.135]{Zaytsev-Thesis2010}, I was able to construct such a specification, which was
subsequently named EDD, for EBNF Dialect Definition. It was then turned into a small nicely packaged paper
for the PL track of SAC~\cite{BNF-WAS-HERE2012} --- the very fact that it was published separately, gave
me a lot of freedom later, when I did not feel like I need to introduce all the metasymbols all over again
in each work that followed.

% ---------------------------------------------------------------------------
\subsubsection{Transforming metasyntaxes}\label{S:XEDD}

\begin{figure*} % htbp
	\centering
		\includegraphics[width=\textwidth]{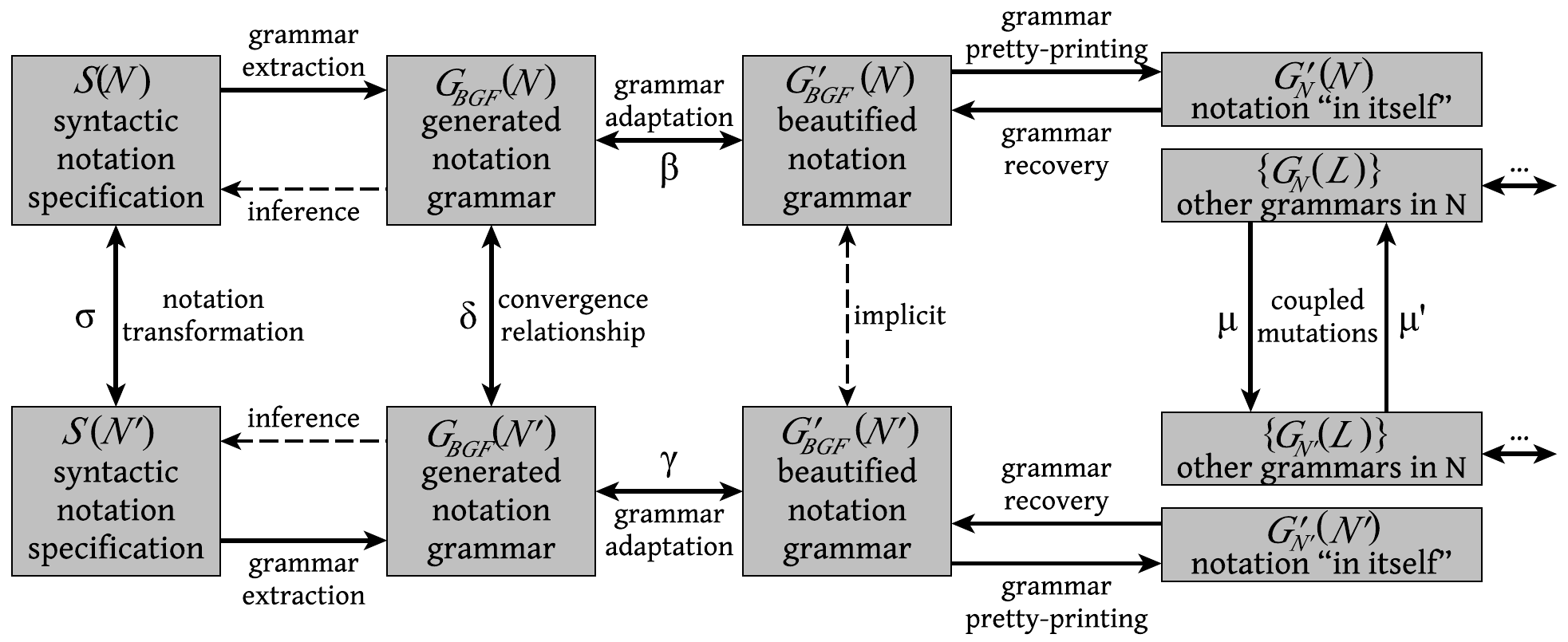}
	\caption{Components of a notation evolution: $\sigma$, a bidirectional
		\emph{notation specification transformation} that changes the notation itself;
		$\delta$, a \emph{convergence relationship} that can transform the notation grammars;
		$\gamma$, a bidirectional \emph{grammar adaptation} that prepares a beautified readable version of
		$N'$.
		$\mu$, an unidirectional \emph{coupled grammar mutation} that migrates the grammarbase according
		to notation changes;
		possibly $\mu'$, an unidirectional \emph{coupled grammar mutation} that migrates the grammarbase
		according to the inverse of the intended
		notation changes.}
	\label{F:bx}
\end{figure*}

Once you have a notation specification as a first class entity, you can define transformations on them. This was probably the first transformation language that I have designed, where the main complexity was not in defining the transformation operators as such, but rather in coupling them with the grammar transformation steps that they imply. The transformation suite consisted of just three operators:

\begin{description}
	\item[rename-metasymbol$(s,v_1,v_2)$]
		{\small where $s$ is the metasymbol and values $v_1$ and $v_2$ are strings}
		\\
		For example, we can decide to update the notation specification from using ``\texttt{:}'' as a
		defining metasymbol to using ``\texttt{::=}''. This is the most trivial transformation, but also
		bidirectional by nature.
	\item[introduce-metasymbol$(s,v)$]
		{\small where $s$ is the metasymbol and $v$ is its desired string value}
		\\
		For example, a syntactic notation can exist without terminator metasymbol, and we may want to
		introduce one.
	\item[eliminate-metasymbol$(s,v)$]
		{\small where $s$ is the metasymbol and $v$ is its current string value}
		\\
		Naturally, eliminate and introduce together form a bidirectional pair. Specifying the current
		value of a metasymbol is not necessary, but enables extra validation, as well as trivial
		bidirectionalisation.
\end{description}

Yet, the final megamodel of the infrastructure that did not even consider language instances (only
grammars and metasyntaxes) looked as complex as \autoref{F:bx}. The paper about evolution of metalanguages
had a bidirectionality flavour and was conditionally accepted at the BX workshop
\cite{Metasyntactically-BX2012}, and then also for the journal special issue \cite{Metasyntactically2012}.

% ---------------------------------------------------------------------------
\subsubsection{Notation-parametric grammar recovery}\label{S:EDDrec}

In all previously published grammar recovery
initiatives~\cite{Recovery-COBOL,Browsable,Recovery-MSC-SSL,Recovery-PLEX,Recovery-SPE,GRK,Too-Sharp2005,Convergence2009,Zaytsev-Thesis2010,JLS-SQJ2011,MediaWiki2011,MediaWiki2012}
the step of transforming the raw grammar-containing text obtained from the language manual was either not
automated (the grammar was re-typed from scratch in the notation required by the target grammarware
framework), or semi-automated (comprised many rounds of test-driven improvement), or automated with a
throwaway tool (one that can not be reused unless the replication deploys exactly the same EBNF dialect).
Having a notation specification as a first class entity, we can step up from throwaway tools to throwaway
notation specifications: at least they take minutes to create, not days.

Notation-parametric grammar recovery~\cite{NPGR-LDTA2012,NPGR2012} was my best result of 2011, and this
year it was officially published and put to several good uses. These uses are not exactly publishable
simply because grammar recovery from (nearly) well-formed has become a trivial process itself, but there
was one story that was enabled by this triviality. The grammar of MediaWiki syntax, for recovery of which
I have a previously exposed preprint~\cite{MediaWiki2011}, is a unique case of using multiple notations
within one community-created grammar. With any other recovery method, it would have been easier to just
retype the grammar again in a uniform fashion, but notation-parametric grammar recovery allowed to treat
all six different incoherent metalanguages with relative ease and derive the final grammar from the
inconsistent input. A continuation of this topic was intended to be a published closure on the case of
MediaWiki grammar recovery, but was unfortunately rejected in the end~\cite{MediaWiki2012}.

% ---------------------------------------------------------------------------
\subsubsection{Notation-driven grammar convergence}\label{S:EDDguided}

Grammar convergence was originally a lightweight verification method not intended for full
automation~\cite{Convergence2009}. However, seeing how many transformations that were in fact converging
grammars, it was possible to infer automatically for the metalanguage evolution case
study~\cite{Metasyntactic2012} (see also \S\ref{S:XEDD}), I could not help starting to wonder whether and
to what extent it was possible to drive the automated convergence process by the notation properties. The
result of that was the methodology of guided grammar convergence, which was already covered by
\S\ref{S:guided}.

% ===========================================================================
\subsection{Tolerance in parsing}\label{S:tolerant}

Originally, the ``parsing in the cloud'' paper \cite{Islands-SCAM2012,Islands-NordiCloud2012} was intended
to present a useful crossing of the in-the-cloud and as-a-service paradigm with the engineering discipline
for grammarware. However, the related work digging quickly got out of hand and turned into a contribution
of its own. The overview of many grammar-based techniques with some level of tolerance towards their input
data and its weak commitment to grammatical structure, was presented at the PEM
Colloquium~\cite{Tolerance2012-talk} (see also \S\ref{S:PEM}), where it was received very warm acceptance
and led to many useful insights. It has been advised to me both by reviewers and colleagues to put more
effort into demonstrative prototype and publish the overview with them separately from the parsing
algorithm (see \S\ref{S:parsing}) itself. This is among one of the planned activities for 2013.

So far, at least the following tolerant parsing methods have been identified:
	ad hoc lexical analysis~\cite{LexicalApproachSucks,AM},
	hierarchical lexical analysis~\cite{Murphy:1995:LSM:222124.222147},
	iterative lexical analysis~\cite{Cox03syntacticapproximation},
	fuzzy parsing~\cite{Fuzzy},
	parsing incomplete sentences~\cite{Lang:1988:PIS:991635.991710}, 
	island grammars~\cite{DocGen},
	lake grammars~\cite{Moonen01}, % island with lakes = parsing incomplete sentences
	robust multilingual parsing~\cite{SynytskyyCD03},
	gap parsing~\cite{GapParsing},
	bridge grammars~\cite{Nilsson-Nyman:2009:PSR:1532448.1532458},
	skeleton grammars~\cite{Tolerant},
	breadth-first parsing~\cite{FortranCompiler,Ophel_breadth-firstparsing},
	grammar relaxation~\cite{DragonBook},
	agile parsing~\cite{DeanCMS03},
	permissive grammars~\cite{KJNV09},
	hierarchical error repair~\cite{SyntaxErrorRepair},
	panic mode~\cite{DragonBook},
	noncorrecting error recovery~\cite{Richter:1985:NSE:3916.4019},
	precise parsing~\cite{Aho:1972:TPT:578789}.
It remains to be seen whether they form a straight spectrum from lexical analysis to strict syntactic analysis.

% ===========================================================================
\subsection{Megamodelling}\label{S:mega}

In computer science, \emph{modelling} happens when a real artefact is represented by its
abstraction, which is then called a model; \emph{metamodelling} happens when the structure of such
models is analysed and expressed as a model for models, or a metamodel; and \emph{megamodelling}
happens when the infrastructure itself, involving multiple models and metamodels, is modelled. The
need for megamodels is being advocated at least since 2004~\cite{Need4Mega,MegaModel}.

The current state of the art is: in the simplest cases, people do not need a special formalism to
state that, for example, ``models A and B conform to the metamodel C''; in somewhat more complicated
scenarios scientists and engineers tend to develop their own domain-specific \emph{ad hoc megamodelling}
methodologies and employ them in narrow domains; and in truly complex situations, any existing
approach only adds to complexity, overwhelming stakeholders with a yet another view on the system
architecture. However, at least one solid business case was found for megamodelling: the problem of
comparing different technological spaces~\cite{KurtevBA02}: for example, comparing the relations
between XML documents, schemata, data models and validators, with relations between object models,
source code and compilers.

At the University of Koblenz-Landau, the Software Languages Team is dedicated to develop a general
purpose megamodelling language called MegaL~\cite{MegaL}. After attending presentations about MegaL
on several occasions, I have paid a working visit to them in July. The consequences of that visit: I
tried to use MegaL for my own megamodelling needs on several
occasions~\cite{Negotiated-XM2012,Negotiated2012,Guided2012}, I have presented an extensive overview
of currently existing ad hoc megamodelling techniques (see \S\ref{S:adhocmega}), and I have proposed
my own method of dealing with overly complex megamodels (see \S\ref{S:renarration}).

% ---------------------------------------------------------------------------
\subsubsection{MegaL dissection}\label{S:adhocmega}

So far at least these previously existing ad hoc megamodelling approaches have been spotted:
	ATL~\cite{Jouault200831},
	UNCOL~\cite{Bratman61,Conway58},
	tombstone~\cite{McKeeman70},
	grammarware megamodelling~\cite{KlintLV05},
	software evolution megamodelling~\cite{MegaModel},
	evolution of software architectures~\cite{Graaf07a,Graaf07b},
	MEGAF~\cite{MEGAF},
	global model management~\cite{VignagaJBB09},
	grammar convergence~\cite{Convergence2009,Zaytsev-Thesis2010,LCI2011},
	software language engineering~\cite{Zaytsev-Thesis2010,SLPS},
	modelling language evolution framework~\cite{Meyers20111223},
	metasyntactic evolution~\cite{Metasyntactically-BX2012,Metasyntactically2012}.

My superficial overview of them, comparing them with MegaL, was presented to the MegaL designers in
July~\cite{MegaL2012-talk}, and my current research activities include active collaboration with them with
a paper presenting a unified model for megamodelling in mind.

% ---------------------------------------------------------------------------
\subsubsection{Renarrating megamodels}\label{S:renarration}

Having seen enough presentations on megamodelling made me realise that they are very easy to follow
even for untrained people, unlike the resulting megamodels that contain far too much detail and are
very intimidating. So, my take on this problem was introducing two operations: slicing (to make
megamodels smaller) and narrating (to traverse the elements in the megamodel). If we have them, we
can take the baseline megamodel that only experts can try to understand, and cut it to consumable
chunks bundled with the story that introduces the remaining elements one by one and explaining each
step. The resulting paper was sent to a workshop on Multiparadigm Modelling, where it was presented
as a poster~\cite{Renarration-MPM2012-talk}, published in online
pre-proceedings~\cite{Renarration-MPM2012} and is currently on its way to the post-proceedings in
the ACM DL~\cite{Renarration2012}.

% ===========================================================================
\subsection{Grammar repository}\label{S:repo}

My first project proposal ever, titled ``Automated Reuse-driven Grammar Restructuring'', was sent to
the NWO Veni program in January, passed a rebuttal phase in May and was finally rejected in July
after informing me that it ended up in the category ``very good''~\cite{NWO-Veni2012}. The idea
described there was small and elegant: mining grammarware repositories. While repository mining
techniques receive quite some attention nowadays, very few people actually have entire repositories
filled with grammars: let's face it, they are omnipresent yet at the same time scarce. However, I
already have this initiative called Grammar Zoo~\cite{SLPS}, which contains many grammars of
languages big and small, and armed with the arsenal of extraction tools developed in my PhD time, it
can grow even more. The goal of such mining is, of course, to reverse engineer reusable grammar
fragments and forward engineer the discipline of their composition.

A paper advocating the need and the usefulness of the repository itself, was written and submitted
to a journal in November~\cite{Zoo2013}. The outcome will only become known in 2013.

% ===========================================================================
\subsection{(Open) Notebook Science}\label{S:ONS}

% presentations \cite{Open2012-talk,Subatomic2012-talk}

Open Notebook Science is an open science paradigm of doing research in a transparent way~\cite{Lloyd2008}.
It involves keeping a lab notebook that collects all data and metadata on experiments, hypotheses,
results, details and other observations that occur during the research phase, so that after the final
objective is reached (or deemed unreachable), the complete path towards it can be exposed and made
publicly available for inspection, replication and reuse. The open notebook approach is fairly well-known
and somewhat popular in fields like biology and chemistry~\cite{DoD2008,Singh2008201}, that strive on
experimental frameworks and traditionally involve lab notebooks, so in practice exercising this approach
had the only consequence of sharing the already existing notebook and systematically referring to it from
the papers. In computer science, however, there are none to few adopters of this approach, mainly due to
the seeming complexity of the method and the amount of extra effort that is needed to set up and to
maintain such a lab notebook and the lack of positive feedback from it in the form of community
encouragement and peer acknowledgement.

During the Software Freedom Day, I have given a presentation, explaining one possible feasible way to
start practicing open notebook science for computer science and software engineering researchers, with the
case study of myself~\cite{Open2012-talk}. A couple of days later SL(E)BOK organisers have heard about it
and asked me to record a keynote presentation~\cite{Subatomic2012-talk} about that, linking open access
ideas with the existing research on ``scientific knowledge objects'' (SKO) and on a ``body of
knowledge''~\cite{Giunchiglia:2010:SKO:2328909.2328928,Liquid2011}.

In short, open notebook science strives to enable open access to atomic SKOs; to expose all the dark
data~\cite{DarkData2007} from failed experiments and unpublished results; to
self-archive~\cite{SelfArchiving} subatomic SKOs, which are relevant for the final result, but smaller
than a ``publon''. Examples of subatomic SKOs include:

\begin{denselista}
	\item Commits to an open source repository;
	\item Tweets on work-related subjects;
	\item Quora answers on work-related topics;
	\item Papers: preprints, reports, drafts, etc;
	\item Presentations: slides, screencasts, etc;
	\item Blog posts;
	\item Wiki edits;
	\item Exposed tools;
	\item Documentation;
	\item Shared raw data;
	\item Auxiliary material.
\end{denselista}

As it has been pointed out to me by some of the attendees of both talks, the topic of subatomic SKOs is
bigger than just open notebook computer science, because if I can show the usefulness of keeping a
notebook of actions for a researcher, it does not necessary mean that the notebook must be public to
profit from its traceability. The first comprehensive paper on this topic is still in the process of being
designed, but hopefully will be submitted somewhere during the next year or two.

%% file: minor.tex
% ===========================================================================
\subsection{Minor topics}\label{S:minortopics}

Additionally to the topics and achievements I consider major for 2012, there are several lesser
contributions: their are either topics that did not receive enough attention to yield a solid major
contribution (yet not insignificant enough to be omitted from the report completely); or just not
traditionally considered worthy of mentioning (programming, engineering, organising effort).

One topic is intentionally hidden from this section, in order to prevent jeopardising an upcoming
submission to a strictly double blind peer reviewed venue.

% ---------------------------------------------------------------------------
\subsubsection{Grammar mutation}\label{S:mutation}

In the paradigm of programmable grammar transformations, the semantics of each of the transformation
operators is bound to the operator itself, and may require arguments to be provided before the actual
input grammar. Such partially evaluated operators (with all arguments provided, but no input grammar yet)
are treated as transformation steps, and their applicability constraints only depend on the grammar: if
they hold, the change takes place; if they do not, an error occurs instead. In other words, the exact
consequence of the transformation step depends on operands, not on the grammar. However, those
applicability constraints can also be processed as filters: whatever part of the grammar satisfies them,
will be transformed --- that way, the exact change in the grammar depends on the grammar, not on the
operands.

As an example, consider renaming grammatical symbols: ``rename nonterminal'' itself is an operator. Its
semantics can be expressed easily on the classic definition of a grammar. If the input grammar is $G =
\langle\mathbb{N},\mathbb{T},\mathbb{P},S\rangle$, then the output must be $$G' =
\langle\mathbb{N}\cap\{x\}\cup\{y\},\mathbb{T},\mathbb{P}|_{x\to y},S'\rangle$$ where $x$ and $y$ are
operands; $S'$ is $S$ unless $S=x$ and $y$ otherwise; and $A|_{x\to y}$ means substitution (for example,
by term rewriting). When $x$ and $y$ are provided, then $G'$ above becomes fully defined and yields
meaningful results when applicability conditions (e.g., $x\in\mathbb{N}$ and $y\not\in\mathbb{N}$) are
satisfied. Renaming a terminal symbol is specified similarly.

However, ``renaming all lowercase nonterminals to uppercase'' is not an operator (or at least even it is
made one, it will be of much higher level than the simple ``rename''), and it is not an atomic
transformation step either: in fact, it can lead to any number of changes in the grammar from $0$ to
$|\mathbb{N}|$, depending on $G$. This number absolutely cannot be known before $G$ is provided.

This kind of grammar manipulation was identified first as a part of research on bidirectional
transformations~\cite{Metasyntactically-BX2012,Metasyntactically2012,BX2012-talk,SEM2012BX-talk} (because
they are not bidirectionalisable), where it received the name of ``grammar mutation''. Later there was an
endeavour to compose a comprehensive list of useful grammar mutations as a part of \cite{Trends2012}, but
it was rejected.

For the sake of providing a better overview of the current state of research on grammar mutations, I
collect all of them in the exhaustive list below. Note that conceptually the same mutations may have been
appearing under different names in various sources: for example, the first mutation in the list, ``remove
all terminal symbols'', has previously been known as a transformation
``\emph{stripTs}''~\cite[\S5.3]{Convergence2009} and as a generator
``\emph{striptxbgf}''~\cite[\S4.9,\S4.10.6.1]{Zaytsev-Thesis2010}.

\begin{description}
	\item[Remove all terminal symbols]
		% striptxbgf, stripTs
		\cite{Convergence2009,Guided-ECMFA2012,Guided-ICSM2012,Guided-POPL2013,Zaytsev-Thesis2010,LCI2011}
		\\
	A simple grammar mutation that is helpful when converging a concrete syntax and an abstract syntax of
	the same intended language. While the abstract syntax definition may have differently ordered
	parameters of some of its constructs, and full convergence will require dealing them them and
	rearranging the structure with (algebraic) semantic-preserving transformations, we will certainly not
	encounter any terminal symbols and can safely employ this mutation.
	\item[Remove all expression selectors]
		% stripsxbgf, stripSs, strips
		\cite{Convergence2009,Guided-ECMFA2012,Guided-ICSM2012,Guided-POPL2013,Zaytsev-Thesis2010}
		\\
	Named (selectable) subexpressions are encountered in many contexts, but the choice of names for them
	is usually even more subjective than the naming convention for the nonterminal symbols.
	\item[Remove all production labels]
		\cite{Guided-ECMFA2012,Guided-ICSM2012,Guided-POPL2013}\hfill\\
	Technically, having production label is the same as making a selectable subexpression out of the right
	hand side of a nonterminal definition. Still, in some frameworks the semantics and/or the intended use
	for labels and for selectors differ.
	\item[Disciplined rename]
		% casexbgf
		\cite{Metasyntactically-BX2012,Metasyntactically2012,Zaytsev-Thesis2010,LCI2011}
		\\
	There are several different well-defined naming conventions for nonterminal symbols in current
	practice of grammarware engineering, in particular concerning multiword names. Enforcing a particular
	naming convention such as making all nonterminal names uppercase or turning camelcased names into
	dash-separated lowercase names, can be specified as a unidirectional grammar mutation (one for each
	convention).
	\item[Reroot to top]
		\cite{LCI2011,Guided-ECMFA2012,Guided-ICSM2012,Guided-POPL2013} \hfill\\
	A top nonterminal is a nonterminal that is defined in the grammar but never used~\cite{Recovery-SPE}.
	In many cases it is realistic to assume that the top nonterminals are intended starting symbols (roots) of the grammar.
	A variation of this mutation was used in \S\ref{S:guided} with an additional requirement that a top nonterminal
	must not be a leaf in the relation graph. This is a rational constraint since a leaf top nonterminal defines a separated
	component.
	\item[Eliminate top]
		\cite{Zaytsev-Thesis2010,LCI2011}
		\\
	In the situations when the root is known with certainly, we can assume all other top (unused) nonterminals to be useless,
	since they are unreachable from the starting symbol and are therefore not a part of the grammar.
	\item[Extract subgrammar]
		\cite{Guided-ECMFA2012,Guided-ICSM2012,Guided-POPL2013} \hfill\\
	Alternatively, we can generalise the last mutation to a parametrised one: given a grammar and a nonterminal (or a list of
	nonterminals), we can always automatically construct another grammar with the given nonterminal(s) as root(s) and the contents
	formed by all production rules of all nonterminals reachable from the assumed root nonterminal(s). Constructing a subgrammar
	starting with the already known roots will eliminate top nonterminals.
	\item[Make all production rules vertical]
		\cite{Zaytsev-Thesis2010,LCI2011,Guided-ECMFA2012,Guided-ICSM2012,Guided-POPL2013} \hfill\\
	Vertical definitions contain several alternative production rules, while horizontal ones have one with
	a top level choice. There are different approaches known to handle this distinction, including
	complete transparency (one form being a syntactic sugar of the other). For normalisation
	purposes or for quick convergence of a consistently vertical grammar and a consistently horizontal
	one, we can use this automated mutation.
	\item[Make all production rules horizontal]
		\cite{Zaytsev-Thesis2010,LCI2011}\hfill\\
	A similar grammar mutation is possible, yet much less useful in practice.
	\item[Distribute all factored definitions]
		\cite{Guided-ECMFA2012,Guided-ICSM2012,Guided-POPL2013}\hfill\\
	Aggressive factoring a-la \texttt{xbgf:distribute} can also be discussed.% in \S\ref{EXBGF}.
	Surfacing all inner choices in a given grammar is a powerful normalisation technique.
	\item[Make all potentially horizontal rules vertical]
		\cite{Zaytsev-Thesis2010,LCI2011}\hfill\\
	Technically, this mutation is a superposition of distribution of all factored definition and converting all resulting horizontal
	production rules to an equivalent vertical form.
	\item[Deyaccify all yaccified nonterminals]
		\cite{Zaytsev-Thesis2010,LCI2011}\hfill\\
	A ``yaccified'' definition~\cite{Adaptation,JongeM01} is named after YACC~\cite{YACC}, a compiler compiler, the old versions
	of which required explicitly defined recursive nonterminals --- i.e., one would write \texttt{A : B} and \texttt{A : A B},
	because in LALR parsers like YACC left recursion was preferred to right recursion (contrary to recursive descent parsers,
	which are unable to process left recursion directly at all). The common good practice is modern grammarware engineering is to use
	iteration metalanguage constructs such as \texttt{B*} for zero or more repetitions and \texttt{B+} for one or more --- this way,
	the compiler compiler can make its own decisions about the particular way of implementation, and will neither crash nor perform
	any transformations behind the scenes. However, many grammars~\cite{SLPS} contain yaccified definitions, and usually the first
	step in any transformation that attempts to reuse such grammars for practical purposes, start with deyaccification, which can
	be easily automated.
	\item[Remove lazy nonterminals]
		\cite{Zaytsev-Thesis2010,LCI2011}\hfill\\
	Many grammars, in particular those that strive for better readability or for generality, contain excessive number of
	nonterminals that are used only once or chain production rules that are unnecessary for parsing and for many other activities
	one can engage in with grammars. We have used an optimising mutation that removes such elements with \emph{xbgf:inline}
	and \emph{xbgf:unchain} on several occasions, including improving readability of automatically generated grammars. 
	% inline or unchain those that are used only once...
	\item[Normalise to ANF]
		\cite{Guided-ECMFA2012,Guided-ICSM2012,Guided-POPL2013}\hfill\\
	The Abstract Normal Form (ANF) was introduced in \S\ref{S:guided} as means of limiting the search space for guided grammar
	convergence. Technically, such normalisation is equivalent to a superposition of removing all labels, removing all selectors,
	removing all terminals, surfacing all inner choices, converting all horizontal production rules to a vertical form, rerooting
	to top non-leaf nonterminals and eliminating others unreachable from them. For conceptual foundations of ANF the reader is
	redirected to the article where it was proposed.
	\item[Fold all grouped subexpressions]
		\cite{Metasyntactically-BX2012,Metasyntactically2012}\hfill\\
	In the context of metalinguistic evolution, we need to construct a coupled mutation for the grammarbase, if the notation change
	contains retiring of a metasyntactic construct that is in use. One of such constructs is the possibility to group symbols
	together in an atomic subsequence --- a feature that is often taken for granted and therefore misused, improperly documented
	or implemented. Naturally, eliminating grouped subexpressions entails folding them to newly introduced nonterminals by means
	of \emph{xbgf:extract}.
	\item[Explicitly encode all separator lists]
		\cite{Metasyntactically-BX2012,Metasyntactically2012}\hfill\\
	Our internal representation of grammars for software languages, following many other syntactic notations, contains a construct
	for defining separator lists. For example: \texttt{\{A ","\}+} is a syntactic sugar for \texttt{A ("," A)*} or
	\texttt{(A ",")* A} --- all three variants specify a comma-separated list of one or more \texttt{A}s. When such a construct
	needs to be retired from the notation, the coupled grammar mutation must refactor its occurrences to explicitly encode separator
	lists with one of the equivalent alternatives.
\end{description} 

A full fledged paper shining enough light on grammar mutations, is still being written and will hit the
submission desks in 2013.

% ---------------------------------------------------------------------------
\subsubsection{Iterative parsing}\label{S:parsing}

As the main (intended) contribution of \cite{Islands-SCAM2012,Islands-NordiCloud2012}, I have proposed the
algorithm for iterative parsing. The basic idea is very simple: we take the baseline grammar and
skeletonise it as far as it can be automated, in such a way that the relation between the ``lakes'' and
the nonterminals in the baseline grammar are preserved. Then, our parse tree will give the basic structure
and a number of watery fragments parsed with useless lake grammars (usually in a form of ``anything but
newline'' or ``something in balanced out curly brackets''). If needed, any of those lakes can be parsed
further with a subgrammar of the baseline grammar, with the new root being the nonterminal that
corresponds to the lake.

This parsing approach was being sold as ``parsing in the cloud'' in
\cite{Islands-SCAM2012,Islands-NordiCloud2012}, which was certainly not the best (even though the coolest)
way to look at it. Other applications for this form of lazy parsing can be found in debugging
(disambiguation, fault localisation) and other areas that traditionally profit from laziness. This remains
future work.

% ---------------------------------------------------------------------------
\subsubsection{Unparsing techniques}\label{S:unparsing}

One of the most confusing paper that I have submitted anywhere in 2012, was the one about unparsing
techniques~\cite{Unparsing-Techniques2012}. Only after finishing writing it, I have realised how big and
overwhelming this topic is. The paper was rightfully rejected after being classified as a ``request for
discussion'': a much deeper survey of (some of) the presented topics must be composed sooner or later, but
it requires much careful consideration. I have not done much in this topic after that, but there was at
least one paper published recently that explicitly considered unparsing~\cite{ObjectGrammars2012}.

The starting idea is simple as a sunrise: there was a lot of effort put in researching parsing techniques,
so why not the opposite? The unparsing techniques can be understood in a very broad sense:
pretty-printing, syntax highlighting, structural import yielding an editable textual representation,
bidirectional construction of equivalent views, etc.

Some papers consider \emph{conservative pretty-printing} as a way to preserve peculiar layout pieces (like
multiple spaces) during unparsing~\cite{PrettyPrint3,Ruckert96}. This is a narrow application of a general
idea of propagating layout through transformations, which is a long-standing and a well-researched
problem. However, even the most conservative unparsers have the risk of introducing an inconsistently
formatted code fragment, if that code was originally introduced by a source code manipulation technique
and not produced by a parser. In other words, replacing a \texttt{GO TO} statement with a \texttt{WHILE}
loop should look differently, depending on how the code around the introduced fragment was formatted.
Possibly, results from the grammar inference research field \cite{Inference2012} can be reused for
recovering formatting rules in some reasonable proximity of the code fragment in order to unparse it
correctly and avoid code alienation.

Suppose not just one desired textual formatted representation of the language instance exists, but several
of them, which form a family, or a product line, like the line of metalanguages considered in
\S\ref{S:meta} in the context of metasyntactic evolution. Following that example, suppose we are given a
grammar in some internal representation and a syntactic notation specification~\cite{BNF-WAS-HERE2012},
then it is somewhat trivial to construct an unparser that would produce the same grammar in a textual
form. In other words, such an unparser should generate a text that, given a notation specification, can
yield the same grammar after automated notation-parametric grammar recovery~\cite[\S3]{NPGR2012}. However,
other questions remain. How to find a minimal notation needed to unparse a given grammar? How in general
to validate compatibility of a given grammar and a given notation? How to produce grammar transformations
(see \S\ref{S:trafo}) to make the grammar fit the notation, how to produce notation transformations (see
\S\ref{S:XEDD}) to make the notation fit the grammar, and how to negotiate to find a properly balanced
outcome? These questions are not trivial and require investigation. Unparser-completeness has recently
been studied in the context of template engines~\cite{AvdBS11}.

Unparsing can also be viewed as commitment to grammatical structure~\cite{KlintLV05}. Can we recover
grammars from them, compare and converge them with other grammars of the same language that we would like
synchronised (e.g., concrete syntax definition intended for parsing, multiple abstract syntaxes for
performing various grammar-based analysis tasks, data models for serialisation)? Are there some specific
properties that such grammars always possess? What is the minimal upper formalism for the baseline grammar
from which grammars for parsing and unparsing can be derived automatically with a language-independent or
language-parametric technology? These questions are not trivial and require investigation.

Connecting to the topic of robust/tolerant parsing (see \S\ref{S:tolerant}), we can consider at least two
kinds of techniques that as the opposite: incremental unparsing and unparsing incomplete trees. By
\emph{incremental unparsing} I mean a modular technique for unparsing modified code fragments and
combining them with the previously unparsed versions of the unmodified code fragments. This is usually not
considered for simple cases, but is possibly worth investigating for large scale scenarios (consider
architectural modifications to an IT portfolio with hundreds of millions lines of code in dozens of
languages). By \emph{unparsing incomplete trees} we define the process of unparsing structured
representations of incomplete language instances. Besides scenarios when this technique is used together
with tolerant/robust parsing (and then the lacking information may be somehow propagated to the unparser
anyway), there are also other scenarios when the gaps are deliberately left out to be filled by the
unparser. In documentation generation, this is the way code examples can be treated --- for a sample
implementation we refer to Rascal Tutor~\cite{RascalTutor}.

For construction of compiler compilers and similar grammarware with unparsing facilities, there is a
commonly encountered problem of bracket minimality for avoiding constructions ambiguous for parsing: since
brackets are there in the text only to guide the parsing process, they are removed from the AST, so how to
put back as few of them as possible during unparsing? This is a typical research question for the
unparsing techniques field. One could also investigate various ways to infer grammar adaptation steps
needed to unparse the given grammatical structure in order to guarantee the lack of ambiguities if it is
to be parsed again.

% \emph{conservative pretty-printing} as a way to preserve peculiar layout pieces (like
% multiple spaces) during unparsing~\cite{PrettyPrint3,Ruckert96}
% 
% inference research field \cite{NierstraszKGLB07,CrepinsekMJBS05,CrepinsekMZ05} 
% 
% Unparser-completeness
% has been studied in the context of template engines in \cite{AvdBS11}.
% 
% rejected \cite{Unparsing-Techniques2012}
% 
% new papers to influence future work:
% 	inference overview \cite{Inference2012}
% and
% 	object grammars \cite{ObjectGrammars2012}
% 
% ---------------------------------------------------------------------------
% \subsubsection{Structured data extraction}
% 
% Wiki Loves Monuments\footnote{\url{http://www.wikilovesmonuments.org}} is a photography contest involving
% volunteers taking pictures of existing state monuments and cultural heritage sites of similar status, and
% releasing them to open access (usually by depositing it to public domain or releasing under a Creative
% Commons license CC-BY or CC-BY-SA). It has been first held in 2010 in the Netherlands. In 2011 
% 
% The existing body of work on matrix grammars and similar grammar variants referenced in the preliminary section, does not consider

% pattern grammars~\cite{grenander1996elements},
% matrix grammars~\cite{SSK72},
% puzzle grammars~\cite{NSSSD91},
% % matrix grammars,
% picture calculus~\cite{MS67}, % picture calculus =?= picture grammars
% picture processing grammars~\cite{Chang:1970:ATP:800161.805166},
% tile grammars~\cite{Reghizzi:2005:TRG:1103398.1103405},
% grid grammars~\cite{YuPaun01},

% ---------------------------------------------------------------------------
\subsubsection{Migration to git}

Following the current trend of leaving old-fashioned open source farms in favour of more modern 2.0 social
coding websites, I have migrated the Software Language Processing Repository from SourceForge to
GitHub~\cite{SLPS}. The project was started in 2008 by Ralf L{\"a}mmel~\cite{FL-is-born} and quickly after
that become the main target for my efforts and the main repository for my code. As of now (December 2012),
it contains 954 revisions committed by me, 314 by Ralf L{\"a}mmel, 44 by Tijs van der Storm and 28 by all
other contributors combined.

This would have not been worth mentioning, if I did not migrate all my other repositories to \texttt{git}
as well, which enabled efficient linking to all of them from the open notebook (see \S\ref{S:ONS}). For
closed source repositories (like ones used for writing papers) we use Atlassian BitBucket instead of
GitHub.

% ---------------------------------------------------------------------------
\subsubsection{Turing machine programming}

Two of my colleagues from Centrum Wiskunde \& Informatica (CWI), Davy Landman and Jeroen van den Bos, have
built a physical Turing machine with a finite tape and separate program space, from LEGO
blocks~\cite{LEGO-Turing-Machine2012}. We were all passively yet encouragingly watching them do that and
then watching with excitement how the resulting machine could sum two and two in less than half an hour.
From the software perspective, they have created a kind of ``Turing assembly'' DSL that consisted of
commands for accessing bits on the tape, moving the head and making decisions on the next command, and was
translatable into some real code that could run on the LEGO chip brick. Then, there was a slightly more
advanced DSL called ``Turing level 2'' developed on top of it, enhanced with label names and repetition
loops, as well as IDE support features like a visualiser/simulator.

My spontaneous contribution to the project involved writing several programs for the machine in this
``Turing language level 2'', including copying of unary numbers, incrementing them, performing various
forms of addition and finally multiplying two unary numbers. All these programs are publicly accessible at the official repository: \url{http://github.com/cwi-swat/TuringLEGO/tree/master/examples}.

% ---------------------------------------------------------------------------
\subsubsection{Grammarware visualisation}\label{S:vis}

Various controversial thoughts on grammar recovery visualisation, related to the previous body
of works on grammar recovery both (co)authored by
me~\cite{Too-Sharp2005,Convergence2009,Zaytsev-Thesis2010,MediaWiki2011,JLS-SQJ2011,BNF-WAS-HERE2012,NPGR2012,MediaWiki2012}
and the giants on shoulders of which I was
standing~\cite{Recovery-COBOL,Browsable,Recovery-MSC-SSL,Recovery-PLEX,Recovery-SPE,KlintLV05},
yielded some experimental code, but no valuable stable results.

In a draft sent to the ``new ideas'' track of FSE 2012~\cite{Grammarware-Visualization2012} to be rejected
there, I have argued that introducing or improving visualisation of processes in grammarware engineering
has at least these benefits:

\begin{description}
	\item[Process comprehension:] it becomes easier to understand the process and to see what exactly is
		happening when it is applied to certain input.
	\item[Process verification:] while complete formal verification of a sophisticated process with many
		branches and underlying algorithms, may be a challenging task, it is relatively easy to pursue
		lightweight verification methods.
		One of them comes more or less for free when an experienced observer can see what is happening and
		detect peculiarities naturally.
	\item[Process improvement:] observing a process does not only let one find mistakes in it, but also to
		get familiar with bottlenecks and other problematic issues, which in turn will help to suggest
		refinements and improvements.
		% we can improve the process itself
	\item[Interactiveness:] there are many examples of processes which are impossible or unfeasibly hard
		to automate completely,
		but for which reasonable automation schemes exist that exercise ``semi-automation'' and require
		occasional feedback from a system operator. The request-response loop for such feedback can be
		drastically shortened in the case of interactive visualisations.
\end{description}

The point of the paper was well-received by the FSE NIER reviewers: nobody tried to argue that
visualisation techniques would be useless. However, I obviously overestimated a contribution that I could
make with providing a ``mile wide, inch deep'' (a quote from one of the reviews) overview, so perhaps a
much later overview with the list of solid achieved results, would be in order. For the sake of
completeness of this report, I list the nine showcases that were briefly described in the NIER submission
below. Each item of this list is a relatively low hanging fruit for an article or a series thereof.

\begin{description}
	\item[Grammar recovery:] the state of the art in automated grammar recovery (see also
	\S\ref{S:EDDrec}) is to work based on a set of appropriate heuristics~\cite{JLS-SQJ2011,NPGR2012}.
	Proper visualisation of them would help: dealing with some particularly tricky notations; verifying
	that the heuristics do what they are intended to do; collecting evidence and statistics on the use of
	certain heuristics; proposing additional heuristics and other process improvements.
	\item[API-fication] is a term used in \cite{KlintLV05} to describe a process of replacing low level
	API calls for manipulating a data structure with more expressive and more maintainable high level API
	calls generated from a grammar~\cite{deJong200435}. Thus, API-fication is a form of grammar-aware
	software renovation where surfacing grammar knowledge is a crucial contribution of the process.
	Visualising both the API calls themselves and the improvement steps on them, can serve as a
	motivation and even as a lightweight verification of API-fication.
	\item[Grammar transformation \& convergence.] There are at least two commonly used ways to visualise a
	grammar: in a textual form as (E)BNF; or as a syntax diagram (``railroad track''). Neither of them has
	a designated visualisation notation for transformations.
	\item[Mapping between grammar notations] is of the biggest challenges in research on grammars in a
	broad sense, since grammarware strives to cover such a big range of various structural definitions.
	Mapping between EBNF dialects~\cite{Metasyntactically2012}, X/O mapping \cite{xotrafo}, O/R mapping
	\cite{ONeil2008}, R/X mapping \cite{Fernandez2002} and many other internotational mappings exist along
	with intranotational techniques for grammar diffing, graph comparison, nonterminal matching, model
	weaving, etc. Displaying matching artefacts in a traceable way by metagrammarware tools is usually
	rather limited and either display local (mis)matches or global statistics.
	\item[Grammarware coevolution.] Concurrent and coupled evolution of grammars and language
	instances~\cite{CicchettiREP08}, of coexisting related grammars~\cite{Adaptation}, of grammars and
	language transformations~\cite{CleveH06}, of language design and implementation~\cite{DHondt2000} are
	special mixed cases of mapping and transformations (see last two sections), where we would like to
	visualise both what kind of matches are made and what kind of actions are inferred from them.
	\item[Grammar-based analysis] comprises syntactic analysis (parsing), but also similarly geared
	techniques that never received enough attention. As an example, it would be great to have something
	to demonstrate hierarchical lexical analysis~\cite{Murphy:1995:LSM:222124.222147} to the same degree
	as \cite{AMUFVI09} demonstrated for LL and LALR parsing.
	\item[Disambiguation] is a process of filtering a parse forest or reasoning about the origins of it,
	in modern generalised parsing algorithms like SGLR~\cite{SGLR} or GLL~\cite{Scott2010177}. Visualising
	SGLR disambiguation~\cite{DisambigSGLR} was implemented in the ASF+SDF Meta-Environment as a part
	of parse tree rendering, so in fact it visualised the ambiguities themselves and not the process of
	removing them, which was still of considerable help. More recent GLL disambiguation
	algorithms~\cite{Basten2010} were expressed mostly in a textual form even within a PhD project
	entirely dedicated to ambiguity detection~\cite{BasPhD} --- primarily because
	there is no clear understanding of how exactly they would be useful to visualise.
	\item[Grammar-based testing] methods based on combinatorial (non-probabilistic) exploration of the
	software language under test, have emerged from recent research~\cite{LaemmelS06,TestMatch2012}.
	Visualising coverage achieved by them and adjusting the visualisation with each new test case should
	help both to keep track of the process by expressing its progress, and to localise grammar fragments
	responsible for the failing test cases.
	\item[Grammar inference] is a family of methods of inferring the grammar, partially or completely,
	from the available codebase and even from code indentation~\cite{CrepinsekMJBS05,NierstraszKGLB07,Inference2012}.
	Such inference is a complicated process based on heuristics and sometimes even on search-based
	methods. As a consequence, each attempt at grammar inference remains somehow unconnected to the rest
	of the research field: adoption of such methods by scientists and engineers outside the original
	working group happens rarely, if ever. One can think that a proper visualisation of such process would
	help new users to get acquainted with a grammar reconstruction system and tweak it to their needs.
\end{description}

\input{jaxb}

NB: the last item was written before the publication of the excellent grammar inference field
overview~\cite{Inference2012}, which can also be seen as considering visualisation in a very broad sense.

% ---------------------------------------------------------------------------

Another newer initiative which can be seen as grammarware process visualisation, concerns guided
convergence (see also \S\ref{S:guided}). We can recall that the whole process of the guided grammar
convergence is rather complicated and involves normalising the input grammar and going through several
phases of unification to ensure the final nonterminal mapping that looks like this~\cite{Guided2012}:

\begin{align*}\mathit{jaxb} \:\diamond\: \mathit{master} =\:& \{\langle \mathit{Expr_2},\mathit{binary}\rangle,\\
 & \langle \mathit{Expr_3},\mathit{conditional}\rangle,\\
 & \langle int,int\rangle,\\
 & \langle \mathit{Function},\mathit{function}\rangle,\\
 & \langle str,str\rangle,\\
 & \langle \mathit{Program},\mathit{program}\rangle,\\
 & \langle \mathit{Expr},\mathit{expression}\rangle,\\
 & \langle \mathit{Expr_1},\mathit{apply}\rangle,\\
 & \langle \mathit{Ops},\mathit{operator}\rangle\}\end{align*}

While preparing the main guided grammar submission, I have noticed that this particular mapping, as well
as the normalised grammar (\autoref{tbl:JAXB}) and the list of weakly and strongly prodsig-equivalent
production rules (\autoref{fig:JAXBvsMG}) can be automatically produced by the convergence tool virtually
without any additional effort in a completely transparent, traceable, reliable and reproducible fashion.
This led to open publication of \cite{Guided2012}, an extended appendix for the main guided grammar
convergence paper, which was, except for the two-page introduction, generated automatically, but is still
readable and useful.

% ---------------------------------------------------------------------------
\subsubsection{Wiki activity}

While contributing to wiki websites is not usually considered an activity worthy of tracking or
mentioning in the academic sense, of the 72 wiki-articles I have written in 2012 I can identify at
least six that can be viewed as (popular) scientific writing:

\begin{denselist}
	\item \href{http://cyclowiki.org/wiki/%D0%93%D1%80%D0%B0%D0%BC%D0%BC%D0%B0%D1%82%D0%B8%D0%BA%D0%B0_%D0%B2_%D1%88%D0%B8%D1%80%D0%BE%D0%BA%D0%BE%D0%BC_%D1%81%D0%BC%D1%8B%D1%81%D0%BB%D0%B5}{Grammar in a broad sense} (11 kB + 1 figure)
	\item \href{http://cyclowiki.org/wiki/%D0%A2%D0%B5%D1%85%D0%BD%D0%BE%D0%BB%D0%BE%D0%B3%D0%B8%D1%87%D0%B5%D1%81%D0%BA%D0%BE%D0%B5_%D0%BF%D1%80%D0%BE%D1%81%D1%82%D1%80%D0%B0%D0%BD%D1%81%D1%82%D0%B2%D0%BE}{Technological space} (16 kB)
	\item \href{http://cyclowiki.org/wiki/%D0%9C%D0%B5%D0%B3%D0%B0%D0%BC%D0%BE%D0%B4%D0%B5%D0%BB%D0%B8%D1%80%D0%BE%D0%B2%D0%B0%D0%BD%D0%B8%D0%B5}{Megamodelling} (6 kB + 2 figures)
	\item \href{http://cyclowiki.org/wiki/%D0%9E%D1%81%D1%82%D1%80%D0%BE%D0%B2%D0%BD%D0%B0%D1%8F_%D0%B3%D1%80%D0%B0%D0%BC%D0%BC%D0%B0%D1%82%D0%B8%D0%BA%D0%B0}{Island grammar} (12 kB)
	\item \href{http://ru.wikipedia.org/wiki/%D0%92%D0%B0%D0%BD_%D0%92%D0%B5%D0%B9%D0%BD%D0%B3%D0%B0%D0%B0%D1%80%D0%B4%D0%B5%D0%BD,_%D0%90%D0%B4%D1%80%D0%B8%D0%B0%D0%BD}{Adriaan van Wijgaarden} (21 kB)
	\item \href{http://ru.wikipedia.org/wiki/%D0%9D%D0%B8%D0%BD%D0%BE%D0%BC%D0%B8%D1%8F,_%D0%A1%D0%BE%D0%BD%D1%82%D0%BE%D0%BA%D1%83}{Ninomiya Sontoku (Kinjiro)} (10 kB)
\end{denselist}

% ---------------------------------------------------------------------------
\subsubsection{Colloquium organisation}\label{S:PEM}

Again, participating in organisation of various events is commonly considered normal for a practicing
academic researcher, but is never counted as a scientific contribution. Not arguing with that, I am still
happy to be able to maintain the existing seminar culture of CWI (Centrum Wiskunde \& Informatica, my
current employer) as a colloquium organiser of a series of events that have been taken place continuously
at least since 1997\footnote{\url{http://event.cwi.nl/pem}}. Over the course of 2012, \textbf{56}
presentations were given in total as a part of Programming Environment Meeting (PEM, mostly an
inter-institutional outlet), Software Engineering Meeting (SEM, mostly an internal group seminar) and a
special one-day event Symposium on Language Composability and Modularity (SLaC'M, most trouble of
organising which was taken by Tijs van der Storm). These speakers have appeared at PEM, SEM and SLaC'M in
2012 (in chronological order of their first appearance):

\newpage\begin{denselist}
	\item \href{http://grammarware.net/}{Dr.~Vadim Zaytsev} \cite{PEM2012-talk,SEM2012BX-talk,Tolerance2012-talk,Replications2012-talk,Decomposition2012-talk}
	\item \href{http://www.cwi.nl/people/2428}{Atze van der Ploeg} \cite{PEM2,PEM34}
	\item \href{http://win.ua.ac.be/~sdemey/}{Prof.~Dr.~Serge Demeyer} \cite{PEM3}
	\item \href{http://www.win.tue.nl/~aserebre/}{Dr.~Alexander Serebrenik} \cite{PEM5}
	\item \href{http://www.linkedin.com/in/stellapachidi}{Stella Pachidi} \cite{PEM6}
	\item \href{http://homepages.cwi.nl/~storm/}{Dr.~Tijs van der Storm} \cite{PEM7,PEM12,SLaCM7}
	\item \href{http://www.informatik.uni-trier.de/~ley/db/indices/a-tree/s/Steindorfer:Michael.html}{Michael Steindorfer} \cite{PEM8,SLaCM5,PEM35}
	\item \href{http://wiki.mq.edu.au/display/plrg/Anthony+Sloane}{Dr.~Antony Sloane} \cite{PEM9}
	\item \href{http://nl.linkedin.com/in/riemervanrozen}{Riemer van Rozen} \cite{PEM10,PEM33}
	\item \href{http://www.cwi.nl/people/2333}{Jeroen van den Bos} \cite{PEM11,PEM21,PEM44}
	\item \href{http://www.linkedin.com/pub/alex-loh/13/971/5a8}{Alex Loh} \cite{PEM13,SLaCM2,PEM31}
	\item \href{http://turingmachine.org/blog/}{Dr.~Daniel M.~German} \cite{PEM14}
	\item \href{http://plg.uwaterloo.ca/~migod/}{Dr.~Michael Godfrey} \cite{PEM15}
	\item \href{http://homepages.cwi.nl/~hills/CWI_Homepage/Homepage.html}{Dr.~Mark Hills} \cite{PEM19,SLaCM4,PEM32}
	\item \href{http://landman-code.blogspot.com/}{Davy Landman} \cite{PEM21,PEM29}
	\item \href{http://www.cwi.nl/people/2528}{Luuk Stevens} \cite{PEM22}
	\item \href{http://gsd.uwaterloo.ca/kczarnec}{Dr.~Krzysztof Czarnecki} \cite{PEM23}
	\item \href{http://www.ii.uib.no/~magne/}{Prof.~Dr.~Magne Haveraaen} \cite{PEM24,PEM25}
	\item \href{http://www.ii.uib.no/~anya/}{Dr.~Anya Helene Bagge} \cite{PEM24,PEM48}
	\item \href{http://homepages.cwi.nl/~simon/}{Dr.~Sunil Simon} \cite{PEM26}
	\item \href{http://www.linkedin.com/in/tbdinesh}{Dr.~T.~B.~Dinesh} \cite{PEM27}
	\item \href{http://homepages.cwi.nl/~jurgenv/}{Dr.~Jurgen Vinju} \cite{PEM28,SLaCM3}
	\item \href{http://www.cs.utexas.edu/~wcook/}{Dr.~William~R.~Cook} \cite{SLaCM1}
	\item \href{http://homepages.cwi.nl/~ai/}{Anastasia Izmaylova} \cite{SLaCM6}
	\item \href{http://www.lclnet.nl/}{Dr.~Lennart Kats} \cite{PEM30}
	\item \href{http://nl.linkedin.com/in/carelbast}{Carel Bast},
		  \href{http://www.wimbast.nl/}{Wim Bast},
		  \href{http://www.linkedin.com/in/tombrus}{Tom Brus} \cite{PEM36}
	\item \href{https://sites.google.com/site/tesfahuntesfay/}{Tesfahun Tesfay} \cite{PEM37}
	\item \href{http://www0.cs.ucl.ac.uk/staff/W.Langdon/}{Dr.~William B.~Langdon} \cite{PEM38}
	\item \href{http://www.linkedin.com/in/andreivaranovich}{Andrei Varanovich} \cite{PEM39}
	\item \href{http://jorisdormans.nl}{Dr.~Joris Dormans} \cite{PEM40}
	\item \href{http://nl.linkedin.com/pub/bas-joosten/3/739/868}{Sebastiaan Joosten} \cite{PEM41}
	\item \href{http://www.linkedin.com/pub/magiel-bruntink/1/4b1/b74}{Dr.~Magiel Bruntink} \cite{PEM42}
	\item \href{http://www.cs.vu.nl/~patricia/Patricia_Lago/Home.html}{Dr.~Patricia Lago} \cite{PEM43}
	\item \href{http://cs.uwaterloo.ca/~ftip/}{Prof.~Dr.~Frank Tip} \cite{PEM45}
	\item \href{http://staff.science.uva.nl/~poss/}{Dr.~Raphael Poss} \cite{PEM46}
	\item \href{http://scherpenisse.net/}{Arjan Scherpenisse} \cite{PEM47}
\end{denselist}

%% file: jaxb.tex
\begin{table}\footnotesize
	\begin{center}
		\begin{tabular}{|l|c|}\hline
		\multicolumn{1}{|>{\columncolor[gray]{.9}}c|}{\footnotesize \textbf{Production rule}} &
		\multicolumn{1}{>{\columncolor[gray]{.9}}c|}{\footnotesize \textbf{Prod. signature}}
		\\\hline
		$\mathrm{p}\left(\text{`'},\mathit{Expr},\mathit{Expr_1}\right)$	&	$\{ \langle \mathit{Expr_1}, 1\rangle\}$\\
		$\mathrm{p}\left(\text{`'},\mathit{Expr},str\right)$	&	$\{ \langle str, 1\rangle\}$\\
		$\mathrm{p}\left(\text{`'},\mathit{Expr},\mathit{Expr_2}\right)$	&	$\{ \langle \mathit{Expr_2}, 1\rangle\}$\\
		$\mathrm{p}\left(\text{`'},\mathit{Expr},\mathit{Expr_3}\right)$	&	$\{ \langle \mathit{Expr_3}, 1\rangle\}$\\
		$\mathrm{p}\left(\text{`'},\mathit{Expr},int\right)$	&	$\{ \langle int, 1\rangle\}$\\
		$\mathrm{p}\left(\text{`'},\mathit{Function},\mathrm{seq}\left(\left[str, \star \left(str\right), \mathit{Expr}\right]\right)\right)$	&	$\{ \langle \mathit{Expr}, 1\rangle, \langle str, 1{*}\rangle\}$\\
		$\mathrm{p}\left(\text{`'},\mathit{Program},\star \left(\mathit{Function}\right)\right)$	&	$\{ \langle \mathit{Function}, {*}\rangle\}$\\
		$\mathrm{p}\left(\text{`'},\mathit{Expr_1},\mathrm{seq}\left(\left[str, \star \left(\mathit{Expr}\right)\right]\right)\right)$	&	$\{ \langle str, 1\rangle, \langle \mathit{Expr}, {*}\rangle\}$\\
		$\mathrm{p}\left(\text{`'},\mathit{Expr_2},\mathrm{seq}\left(\left[\mathit{Ops}, \mathit{Expr}, \mathit{Expr}\right]\right)\right)$	&	$\{ \langle \mathit{Ops}, 1\rangle, \langle \mathit{Expr}, 11\rangle\}$\\
		$\mathrm{p}\left(\text{`'},\mathit{Expr_3},\mathrm{seq}\left(\left[\mathit{Expr}, \mathit{Expr}, \mathit{Expr}\right]\right)\right)$	&	$\{ \langle \mathit{Expr}, 111\rangle\}$\\
		\hline\end{tabular}
	\end{center}
	\caption{The JAXB grammar in a broad sense: in fact, an object model obtained by a data binding
	framework. Generated automatically by JAXB~\cite{JSR31} from the XML schema for the Factorial Language~\cite{Guided2012}.}
	\label{tbl:JAXB}
\end{table}

\begin{figure*}
	\begin{eqnarray*}
	\mathrm{p}\left(\text{`'},\mathit{Expr},\mathit{Expr_1}\right) & \bumpeq & \mathrm{p}\left(\text{`'},\mathit{expression},\mathit{apply}\right) \\
	\mathrm{p}\left(\text{`'},\mathit{Expr},str\right) & \bumpeq & \mathrm{p}\left(\text{`'},\mathit{expression},str\right) \\
	\mathrm{p}\left(\text{`'},\mathit{Expr},\mathit{Expr_2}\right) & \bumpeq & \mathrm{p}\left(\text{`'},\mathit{expression},\mathit{binary}\right) \\
	\mathrm{p}\left(\text{`'},\mathit{Expr},\mathit{Expr_3}\right) & \bumpeq & \mathrm{p}\left(\text{`'},\mathit{expression},\mathit{conditional}\right) \\
	\mathrm{p}\left(\text{`'},\mathit{Expr},int\right) & \bumpeq & \mathrm{p}\left(\text{`'},\mathit{expression},int\right) \\
	\mathrm{p}\left(\text{`'},\mathit{Function},\mathrm{seq}\left(\left[str, \star \left(str\right), \mathit{Expr}\right]\right)\right) & \Bumpeq & \mathrm{p}\left(\text{`'},\mathit{function},\mathrm{seq}\left(\left[str, \plus \left(str\right), \mathit{expression}\right]\right)\right) \\
	\mathrm{p}\left(\text{`'},\mathit{Program},\star \left(\mathit{Function}\right)\right) & \Bumpeq & \mathrm{p}\left(\text{`'},\mathit{program},\plus \left(\mathit{function}\right)\right) \\
	\mathrm{p}\left(\text{`'},\mathit{Expr_1},\mathrm{seq}\left(\left[str, \star \left(\mathit{Expr}\right)\right]\right)\right) & \Bumpeq & \mathrm{p}\left(\text{`'},\mathit{apply},\mathrm{seq}\left(\left[str, \plus \left(\mathit{expression}\right)\right]\right)\right) \\
	\mathrm{p}\left(\text{`'},\mathit{Expr_2},\mathrm{seq}\left(\left[\mathit{Ops}, \mathit{Expr}, \mathit{Expr}\right]\right)\right) & \Bumpeq & \mathrm{p}\left(\text{`'},\mathit{binary},\mathrm{seq}\left(\left[\mathit{expression}, \mathit{operator}, \mathit{expression}\right]\right)\right) \\
	\mathrm{p}\left(\text{`'},\mathit{Expr_3},\mathrm{seq}\left(\left[\mathit{Expr}, \mathit{Expr}, \mathit{Expr}\right]\right)\right) & \bumpeq & \mathrm{p}\left(\text{`'},\mathit{conditional},\mathrm{seq}\left(\left[\mathit{expression}, \mathit{expression}, \mathit{expression}\right]\right)\right) \\
	\end{eqnarray*}	\caption{Matching of production rules with the Abstract Normal Form of the JAXB-produced grammar on the left and the master grammar on the right~\cite{Guided2012}.}
	\label{fig:JAXBvsMG}
\end{figure*}

%% file: venues.tex
\section{Venues}\label{S:venues}

% As we all find out sooner or later, different
Academic venues (mostly conferences, workshops and journals) are essential components of the research
process: publishing there means community recognition; submitting eventually leads to receiving peer
reviews; and even reading calls for papers can be very inspiring and eye-opening. Below I list two kinds
of venues that contributed to my research in 2012: one list is for those where I have submitted, the other
one for the rest --- I am deeply grateful to all the reviewers and organisers of both kinds. The lists are
not meant to cover all possible venues for my field, just those directly relevant to my activities this
year.

\subsection{Exercised venues}\label{S:yesvenues}

\begin{description}
	\item[BX 2012 (ETAPS workshop)]~\\
		I have been a ``\emph{bx-curious}'' person for quite a while, but BX 2012 was my first venue
		to come out.
		A very inspiring call for papers\footnote{\url{http://www.program-transformation.org/BX12}},
		excellent atmosphere during the workshop, friendly and productive reviewers. A typical
		example of an event that appreciates you preparing a dedicated paper for which this becomes
		the one and only target venue. I submitted against all the odds (December deadlines are
		rather stressful), got there against all the odds (had to fly from ETAPS to SAC and then
		back) and still regretted nothing. I will not attend BX 2013 (my grandmother has her 80th
		birthday on the day of the workshop, and one has to set priorities), but I would if I could.
		Definitely recommended for people at least marginally interested in this field~\cite{bxReport}.
	\item[SAC 2012 (PL track)]~\\
		A yet another experimental submission in the sense that I did not know almost anyone from
		the programme committee at that moment. However, I know people from my technological space
		who published there, and the
		call for papers\footnote{\url{http://www.cis.uab.edu/bryant/sac2012}}
		was inspiring, so I gave it a try, and did not regret it. The whole conference is huge, so
		I was afraid that attending would be unproductive, but I was proven wrong: if you know at
		least a couple of people with similar research interests and stick to them all the time,
		you will find many other similar researchers to talk to. I did not submit anything to SAC
		2013 due to bad planning (holidays right before the deadline are unproductive), but
		I definitely will consider it very seriously every year from now on.
	\item[LDTA 2012 (ETAPS worshop)]~\\
		Trying to be a good programme committee member, I knew I have to attend, so I have submitted
		the best result of 2011 there: the Grammar Hunter. I was also pleased to see how the current
		call for papers\footnote{\url{http://ldta.info/ldta_2012_cfp.pdf}} positioned LDTA as ``SLE, but
		with more grammarware''. The future of LDTA remains to be determined, but it has departed from
		ETAPS and will most probably join forces with SLE.
	\item[ECMFA 2012]~\\
		The call for papers\footnote{\url{http://www2.imm.dtu.dk/conferences/ECMFA-2012/contributions/?page=cfp}}
		made it look like I have a chance, so I submitted something that I believed to be of good quality
		and of possible interest to the modelware researchers. One of the reviewers said that the paper
		``clearly makes the most contribution of any paper I read'', which was rather encouraging, but
		ended up with rejection. In the end, I must conclude that I should have devoted this time to
		writing for ICPC or one of the journal special issues with deadlines around early spring.
	\item[TFP 2012]~\\
		The call for papers\footnote{\url{http://www-fp.cs.st-andrews.ac.uk/tifp/TFP2012/TFP_2012/CFP12.txt}} looked
		challenging, but I really liked the ``trends'' aspect of it, since most traditional
		conferences dislike overview papers unless they are extremely strong and retrospective:
		there is simply no place for overviews of the current trends, unless you are already in the
		field and you systematically explore the ``future work'' sections of all papers you come by.
		In contrast to BX, this was an example of a venue that did not appreciate preparing a
		paper specifically for them on a topic relevant to me. In less than two weeks after
		submission I have received a short notification that it was judged to be out of scope.
		This was obviously not the only reason since other (stronger, less ``trendy'') papers from my technological
		space like \cite{DBLP:journals/corr/abs-1201-0024} were accepted, so I can only conclude
		that I have failed to explain the link between grammar transformation and the functional
		programming paradigm properly. Given the fact that I am not qualified to report on
		``trends'' in any other field, I doubt that I will try sending anything to this venue in
		the future, but I surely do not discourage others to do so. Personally for me, it would have been
		more more productive to pursue MoDELS which had a competing deadline this year.
	\item[JUCS (journal)]~\\
		The call for papers\footnote{\url{http://www.jucs.org/ujs/jucs/info/special_issues/sbcars_cfp.pdf}} made it clear that
		this special issue is linked to a workshop where I did not participate, but the call was open, and
		I answered. I cannot say that that was very appreciated: the reviews for \cite{MediaWiki2012} came
		very late (several months after the notification deadline), were extremely short and discouraging.
	\item[SCAM 2012]~\\
		This is the third time I have served as a programme committee member for SCAM, where I have
		been invited after our paper with Ralf L{\"a}mmel got a best paper award in
		2009~\cite{JLS-SCAM2009}. I have never attended since that time, and received a warning that
		I will not be included next year if I miss the event again. So, putting date-conflicting
		events like SLE and CSMR aside, I did my best, which for me meant submitting one paper to
		SCAM and one to the colocated ICSM (see below). The topic chosen for SCAM (island grammars)
		seemed to be in scope of the call for papers\footnote{\url{http://scam2012.cs.usask.ca/CFP}},
		but the paper was seen as weird and immature, and was hopelessly rejected. The reviews it
		received were pretty helpful, even though one of the reviewers really hated the ``in the
		cloud'' aspect (and that is exactly how I tried to sell it). Apparently, putting some effort into
		submitting something has already been noticed, since I have been, against all the odds, invited
		to the programme committee again for SCAM 2013.
	\item[ICSM 2012]~\\
		The call for papers\footnote{\url{http://selab.fbk.eu/icsm2012/download/cfp-icsm2012.pdf}} came
		to my attention right after the rejection letter from ECMFA, and I decided that ICSM would be a
		good venue for the guided grammar convergence methodology (\S\ref{S:guided}). Getting a paper
		there would also increase my chances at going to SCAM (see above for the reasons). Reviews were
		rather cold, but some of them (except one) useful nonetheless.
	\item[NordiCloud (WICSA/ECSA workshop)]~\\
		Not really being an architecture researcher,
		I would have never considered going to WICSA/ECSA, but the call for
		contributions\footnote{\url{http://46.22.129.68/NordiCloud/?page_id=39}} was out precisely a
		couple of weeks after my SCAM rejection, and I was not feeling enough energy to rewrite the island
		parsing paper completely, so NordiCloud was a relatively cheap way for me to resubmit the same
		material after a minor revision. It did not pay off: most of the reviewers were scared off just
		by seeing a grammar-related submission.
	\item[FSE 2012 (NIER track)]~\\
		The call for papers\footnote{\url{http://www.sigsoft.org/fse20/cfpNewIdeas.html}} came out at a
		very busy time, but four page limit was easily reachable, so I have submitted two papers on
		different new ideas. Unfortunately, they were indeed more of idealistic proposals for discussing
		and considering certain aspects, than usual ``short papers'' that are just normal papers at the
		early stage. Both were hopelessly rejected, and I still want to find some venue for the future
		that would be good for sharing and discussing fresh ideas --- perhaps OBT? I have to try to find
		out.
	\item[SoTeSoLa 2012 (summer school)]~\\
		An experiment in ``Research 2.0'' driven mostly by Jean-Marie Favre and Ralf L{\"a}mmel, this
		summer school was by far not a typical one. There was a lot of innovations: submitting a one-page
		profile of yourself, making a one-minute video about yourself, listening to lots of remote
		lectures, having a hackathon distributed in time and space, registering at a social networking
		website, etc. Not all of them very entirely successful: partly due to being ahead of its time,
		partly due to other reasons, which are being dissected, analysed and researched now by Jean-Marie
		Favre. I was involved in all kinds of activities from the relatively early
		stage, and in the end it was officially classified as serving as a ``Social Media Chair'' and a
		``Hackathon Lead Coordinator''. This was not a publishing venue, and I did not give any invited
		lecture, but it was fun to be a part of it.
	\item[SATToSE 2012 (seminar)]~\\
		A non-publishing seminar series where I have given a presentation on bidirectional grammar
		transformation~\cite{BGX2012-talk}. The material presented there was in a state somewhere between
		\cite{Metasyntactically2012} and the planned future paper on bidirectionalisation.
	\item[POPL 2013]~\\
		The call for papers\footnote{\url{http://popl.mpi-sws.org/2013/popl2013-cfp.pdf}} was concise and
		crunchy, but POPL is one of the venues that does not require much advertisement. I have poured
		a lot of effort into \cite{Guided-POPL2013}, completely redesigned the convergence process (see
		\S\ref{S:guided}), reimplemented the prototype and rewritten the paper with respect to
		\cite{Guided-ECMFA2012,Guided-ICSM2012}. In a way, it did pay off: the paper was rejected, but
		the reviews were among the most useful that I have received this year.
	\item[NWPT 2012]~\\
		The call for papers\footnote{\url{http://nwpt12.ii.uib.no/call-for-papers}} was brought to my
		attention by Anya Helene Bagge, a co-organiser of this workshop. In an extended abstract that was
		submitted there, I apparently went overboard with the required abstraction level and assumed level
		of grammatical knowledge, and recent POPL rejection has possibly jeopardised the outsourcing of
		the usefulness statement of the method. Reviews were curt and bleak.
	\item[EMSE (journal)]~\\
		The call for papers\footnote{\url{http://sequoia.cs.byu.edu/lab/?page=reser2013&section=emseSpecialIssue}} called for
		``experimental replications'' and went to great length explaining how important it is to be able
		to publish not just the experiments themselves, but also replications thereof. I was immediately
		convinced, but decided to reinterpret the definition of a replication. Instead of doing classical
		empirical studies, I presented research activities (and in particular prototype engineering)
		as experiments. That way, the replications were also ``experiments'' in that sense that were
		intended to cover an older experiment and could therefore be measured and assessed based on
		grounds of that coverage. I could even find some related work on the topic in form of papers that
		described the prototype development process itself.
		My paper was intended to contain three case studies:
		(1) replicating the grammar convergence case study of the Factorial Language from
		\cite{Convergence2009} with the guided grammar convergence methodology (see \S\ref{S:guided});
		(2) replicating a bigger grammar convergence case study of Java from \cite{JLS-SQJ2011}
		with more abstract and concise Extended XBGF (see \S\ref{S:EXBGF});
		(3) replicating both of these case studies with a bidirectional \CBGF\ (see \S\ref{S:CBGF}).
		Due to insane amounts of work that this turned out to be, only the first two replications have
		made it into a 42-page long paper~\cite{Incremental2012}. Only one of three reviewers was excited
		by my approach, and all three agreed that the empirical software engineering journal is not the
		right venue for such a report.
	\item[MPM 2012 (MoDELS worshop)]~\\
		Basically, this venue was chosen \emph{after} I have written a paper. The text underwent
		some polishing after the choice was made, but the topic was not adjusted. I have had a nice idea
		of transforming megamodels in order to make a good story out of them (see \S\ref{S:mega}): a
		substantial contribution was not yet there (and such work is still ongoing), but I wanted to
		expose it to the public and to discuss it first. 
		The call for papers\footnote{\url{http://avalon.aut.bme.hu/mpm12/MPM12-CFP.pdf}} for MPM looked
		the most inviting for this kind of cross-paradigm approach among all MoDELS workshops, and indeed
		the reviewers found the paper weird yet acceptable, so I was able to give a short presentation
		and hang my poster there~\cite{Renarration-MPM2012-talk}.
	\item[XM 2012 (MoDELS worshop)]~\\
		The topics list\footnote{\url{http://www.di.univaq.it/XM2012/page.php?page_id=13}} provided by the
		organisers of this workshop was fascinating, and I desperately wanted to submit anything, but
		eventually gave up to find the time it deserved. Soon after that, the deadline was extended, and I
		had no other choice than to write down the idea that was floating around in my head for
		a while (see \S\ref{S:NBGF}).
	\item[SCP (journal)]~\\
		The call for systems\footnote{\url{http://www.win.tue.nl/~mvdbrand/SCP-EST}} was very much in
		sync with what its guest editors have tried to achieve in the last years, and I support them
		wholeheartedly in that. The Grammar Zoo, one of essential parts of the SLPS~\cite{SLPS},
		that did not receive a lot of my attention in 2012, but that was always on my mind, was packaged
		and submitted there both as an available system and as an important repository of experimental
		systems in grammarware. The outcome will become known in 2013.
\end{description}

% ===========================================================================
\subsection{Inspiring venues}\label{S:novenues}

There have been many venues that I did not submit anything to, but not because I did not want to. Their
calls for papers gave me inspiration to work on something, even though I was not productive enough to be
able to fit into their deadlines or produce anything of value at the required level.

\begin{description}
	\item[MSR 2012]~\\
		The mining challenge\footnote{\url{http://2012.msrconf.org/challenge.php}} of MSR looked very
		interesting, so I looked at it, but since I was looking specifically for grammars, it did not
		work out at all: only two grammars were found, and there was no sensible way to connect them to
		the rest of the system. If more of them could have been obtained written in a variety of EBNF
		dialects, it could have become an interesting case study similar to \cite{MediaWiki2012}.
	\item[Laws and Principles of Software Evolution]~\\
		The call for papers\footnote{\url{http://listserv.acm.org/scripts/wa-acmlpx.exe?A2=ind1111&L=seworld&F=&S=&P=25841}}
		for this special issue of JSME looked tempting, so I even emailed the editors, asking for some
		additional information. Unfortunately, the collaboration that I hoped to achieve with other
		people, did not work out, and nothing was produced in time.
	\item[Success Stories in Model Driven Engineering]~\\
		The call for papers\footnote{\url{http://www.di.univaq.it/ssmde}} came out at the time when I was
		busy with all kinds of other initiatives. Besides that, this special issue of SCP was actually
		looking for extended reports on already published projects, and I was busy with new experiments.
		Possibly, a strong ``lessons learnt'' kind of paper on grammar hunting would make sense, but I was
		too immersed in new stuff at the time to go back. However, I have to admit that when/if I finally
		sit down to write a comprehensive grammar recovery paper (i.e., connecting \S\S\ref{S:EDDrec},
		\ref{S:mutation}, \ref{S:vis} and \ref{S:repo}), it must go to either SCP or SP\&E.
	\item[CloudMDE (ECMFA 2012 workshop)]~\\
		This was \emph{the} venue that gave me the eerie thought of writing a ``parsing in the cloud''
		paper (see \S\ref{S:parsing}). However, I was disheartened by the rejection of
		\cite{Guided-ECMFA2012} at ECMFA and decided to not submit anything to ECMFA
		workshops\footnote{V.~Zaytsev (grammarware). ``Yet another bridging attempt failed: my grammar
		paper got rejected at @\href{http://twitter.com/ecmfa2012}{ecmfa2012}. Now I will also go submit
		the \#CloudMDE draft elsewhere.'' Tweet. \url{https://twitter.com/grammarware/status/189976445995593728}. 11 April 2012, 9:21.}.
	\item[ICPC 2012]~\\
		The call for papers\footnote{\url{http://icpc12.sosy-lab.org/CfP.pdf}} competes date-wise with
		many other good venues, so this year ICPC just happened to not be among the ones I have chosen
		as my targets.
	\item[RC 2012]~\\
		The call for papers\footnote{\url{http://www.reversible-computation.org/2012/index91b1.html?call_for_papers}}
		for the fourth workshop on reversible computation gave me a lot of ideas and keyword pointers
		for the bidirectionality topic. However, I did not feel confident enough to submit anything.
		Anyway, thanks a lot and congratulations on becoming a conference in 2013!
	\item[CSCW 2013]~\\
		This is not a typical venue for me, but I have a dream of eventually submitting something
		wiki-related there.
		The call for participation\footnote{\url{http://cscw.acm.org/participation_paper.html}} was
		as good as it always is, and even better this year because they have introduced a new rule
		concerning the paper size: 10 pages is no longer the \emph{limit}, it is rather a \emph{standard}.
		If your idea fits on smaller number of pages, your reviewers have the right to complain if you
		try to bloat your submission. On the other hand, if that is not enough, you can always make your
		paper longer, but the contribution then needs to grow accordingly.
		I believe that with small incremental and non-disruptive ideas like these, we could achieve
		modern comfortable publishing models easier than with endeavours to revolutionise the field.
	\item[PPDP/LOPSTR 2012]~\\
		The calls for papers\footnote{\url{http://dtai.cs.kuleuven.be/events/PPDP2012/ppdp-cfp.txt}}$^,$\footnote{\url{http://costa.ls.fi.upm.es/lopstr12/cfp.pdf}} were both interesting, but at my level I could not actually decide
		between the two venues. I was working honestly toward the seemingly achievable goal (see
		\S\ref{S:3BGF}), but it turned out to be unachievable. Being insecure about my ability to write
		a strong paper about negative results, I gave up.
	\item[FM+AM 2012 (SEFM workshop)]~\\
		This was \emph{the} workshop\footnote{\url{http://ssfm.cs.up.ac.za/workshop/FMAM12.htm}} that
		set my thoughts in the agile/extreme mode, which ultimately led to the paper at XM (see
		\S\ref{S:NBGF}) simply because I did not manage to complete the work before the FM+AM deadline.
		Imagine my surprise when I found out that FM+AM was cancelled due to the lack of good submissions!
	\item[WoSQ 2012 (FSE workshop)]~\\
		The call for papers\footnote{\url{http://sites.google.com/site/wosq2012/cfp}} has led me to
		believe that this would be a good possible venue for the paper on grammar mutations (see
		\S\ref{S:mutation}). However, the time was too tight, and both of my NIER submissions have been
		rejected, so an FSE workshop stopped looking that attractive after all.
	\item[SQM 2012 (CSMR workshop)]~\\
		The workshop\footnote{\url{http://sqm2012.sig.eu}} happened at the same time as I was attending
		both ETAPS and SAC, so I could not possible be at the third place at the same time as well, but
		I just want to name it as a relatively small venue where I have enjoyed reviewing a couple of
		papers as a PC member (will be on PC next year as well).
	\item[WCN 2012]~\\
		The website\footnote{\url{http://www.wikimediaconferentie.nl}} is in Dutch, as the conference
		itself. This was my second experience being a Program Chair (the first one was with WCN 2011),
		and this time I counted: 842 emails needed to be sent or answered by me in order for this
		conference to happen. Luckily, CWI (my employer) did not mind since they could proudly list
		``one of theirs'' to be the PC at a venue where one of the keynote speakers is Jimmy
		Wales~\cite{JimboWCN12}.
\end{description}

% Veni: 5 Jan as a deadline is crazy
% TOOLS, iFM, CAV, ICMT, XP — deadlines too early in the year
% MSR: tried the challenge, didn’t find anything interesting wrt grammars
% JSME: Laws and Princinples of Software Evolution — missed
% SCP: Success Stories in Model Driven Engineering — missed
% ECMFA: frankly, invested effort did not pay off
% should have gone for ICPC, JSME, SCP…
% TFP: same, the effort did not pay off
% should have pursued MODELS instead
% BX: very cool, true workshop
% RC: inspiring CfP in the context of BX
% LDTA: tough discussion, I’m a PC Chair next year
% CSCW: very inspiring, needed more time
% PPDP: tried, failed because the results were tough to interpret
% FM+AM: inspiring CfP
% WoSQ
% SQM: thanks for asking for being in a PC
% SCAM: thanks for all the fish
% WCN: PC Chair + Jimmy Wales

%% file: conclusion.tex
\section{Concluding remarks}\label{S:end}

\subsection{Immediate results}

Writing this extended year report has achieved at least three goals:

\begin{description}
	\item[Streamlining new ideas.] 
	Reexplaning (renarrating?) research ideas and putting them in perspective
	has helped to crystallise them into publishable achievable objectives.
	\item[Knowledge dissemination.]
	This document can serve both as a scientific report for my colleagues and
	superiors, and as an entrance point for people who want to get acquainted with my results for other
	reasons.
	\item[Case study in self-archiving.]
	As I have said in the introduction, this report can be seen as advanced form of self-archiving.
	It was a relatively big effort, compared to the traditional ``just put the PDF online'' thing.
	Together with the open notebook initiative, it stressed the paradigm and raised some questions yet
	to be answered (e.g.: How to properly break one atomic SKO --- essentially, a publication --- into
	subatomic ones to distinguish ``I've done for the tool that was later described in this paper'' from
	``I've done this for the particular version of this paper, which was later rejected''? What are all
	possible stages in the SKO lifecycle?).
\end{description}

% 
% two shades of green: ``done done''; ``almost done, pending DOI'' --- to fully comply with ACM copyright policy, the author's version of the paper distributed for self-archiving purposes, must contain the DOI.

\subsection{Special features}

\begin{description}
	\item[The presence of an open notebook.] 
	A lot of claims about dates, continuations and amounts of effort, made on the pages of this report, can
	be reformulated into queries on the open notebook, and formally validated as such. For now these claims
	were intentionally done in plain text because no reliable or traditionally acceptable infrastructure
	exists for them (yet).
	\item[Open access window.]
	All the papers mentioned here, were put online immediately after their submission (unless prohibited
	explicitly by the submission rules), and taken down immediately after their rejection (if any). At this
	stage, I do not know any better way to expose your research results to the public: official acceptance
	can take months and years, during which one could have profited from sharing the contents around.
	\item[Rejected material.]
	Not all rejected papers are rejected because they are inherently, irreparably bad: some turn out to be
	out of scope, lacking some essential results or simply not mature enough to be published (yet). With
	this report, I have exposed most of the dark data concerning my rejected material.
	\item[Unfruitful attempts.]
	Also classified as dark data by \cite{DarkData2007}, but of an entirely different nature: these are
	failed experiments: prototypes that have never made it to the point of being ready to be described in
	a paper. There can be traces of such unfruitful attempts in presentations and other subatomic SKOs
	before their futility becomes apparent.
	\item[Venues.]
	Knowledge about workshops, conferences and journals seems to float around in the academic community
	and is usually distributed as folklore, if at all. There are many reasons for doing so, ranging from
	the lack of incentive to the fear of occasional offence.
\end{description}

% 
% timely open access window to active drafts
% 
% rejections \cite{DarkData2007}
% 
% unfruitful attempts
% 
% venues

\subsection{Acknowledgements}

I am most grateful to the following people, without whom 2012 would not have been the same:

\begin{denselist}
	\item Ronald de Wolf, Bas de Lange, Tijs van der Storm, Jean-Marie Favre --- for inviting me to present at various events.
	\item Ralf L{\"a}mmel --- for inviting me for a working visit to Koblenz and hosting me there.
% \end{denselist}\newpage\begin{denselist}
	\item Felienne Hermans --- for referring me to the matrix grammars, picture calculus and adjacent topics.
	\item Paul Klint, Jurgen Vinju, Tijs van der Storm, Sander Bohte --- for contributing to organising the PEM Colloquium.
	\item T.~B.~Dinesh --- for introducing me to the notion of renarration.
	\item Frans Grijzenhout and Sandra Rientjes --- for coorganising a conference with me.
	\item Lodewijk Gelauff --- for introducing me to the topic of structured data extraction.
	\item Mark van den Brand --- for appreciating the input in the discussion on the future of LDTA, and for putting trust in me for chairing its programme next year.
	\item Jurgen Vinju, Tijs van der Storm --- for collaborating on a grammar-related topic.
	\item Ralf L{\"a}mmel, Andrei Varanovich, Jean-Marie Favre --- for collaborating on megamodelling topics.
	\item Zinovy Diskin --- for insightful discussions during ETAPS'12 and MoDELS'12 on category theory and grammars in a broad sense.
	\item Anonymous reviewers of BX'12, SAC'12, LDTA'12, ECMFA'12, ICSM'12, SCAM'12, NordiCloud'12, FSE NIER'12, XM'12, MPM'12, NWPT'12, POPL'13, JUCS, EMSE and NWO Veni programme --- for the effort that they have put into considering, assessing and reviewing my work.
	\item All presenters at PEM and WCN --- for their contributions, the only essential component of the final product.
	\item All unmentioned colleagues and uncounted conference contacts --- for fruitful discussions.
	\item All people who tried to contact me through email --- for patience and unreasonable waiting times.
\end{denselist}